\documentclass[fleqn,usenatbib]{mnras}

\usepackage[T1]{fontenc}
\usepackage{ae,aecompl}

\usepackage{graphicx}	
\usepackage{amsmath}	
\usepackage{amssymb}	
\usepackage{newtxtext,newtxmath}

\title[The PAU survey: Ly$\alpha$ intensity mapping forecast]{The PAU survey: Ly$\alpha$ intensity mapping forecast}

\author[P. Renard et al.]{
Pablo Renard$^{1,2}$\thanks{E-mail: p.renard.guiral@gmail.com},
Enrique Gaztanaga$^{1,2}$,
Rupert Croft$^{3}$,
Laura Cabayol$^{4}$,
\newauthor
Jorge Carretero$^{4,5}$,
Martin Eriksen$^{4}$,
Enrique Fernandez$^{4,6}$,
Juan Garc\'{\i}a-Bellido$^{7}$,
\newauthor
Ramon Miquel$^{4,8}$,
Cristobal Padilla$^{4}$,
Eusebio Sanchez$^{9}$
and Pau Tallada-Cresp\'{i}$^{5,9}$
\\
$^1$ Institute of Space Sciences (ICE, CSIC), Campus UAB, Carrer de Can Magrans, s/n,  08193 Barcelona, Spain\\
$^2$ Institut d'Estudis Espacials de Catalunya (IEEC), E-08034 Barcelona, Spain  \\
$^{3}$Department of Physics, Carnegie Mellon University, 5000 Forbes Avenue, Pittsburgh 15213, United States\\
$^4$ Institut de F\'{i}sica d'Altes Energies (IFAE), The Barcelona Institute of Science and Technology, Campus UAB, 08193 Bellaterra (Barcelona), Spain\\
$^5$ Port d'Informaci\'{o} Cient\'{i}fica (PIC), Campus UAB, C. Albareda s/n, 08193 Bellaterra (Barcelona), Spain \\
$^6$ Universitat Aut\'{o}noma de Barcelona, Bellaterra, Barcelona, Spain\\
$^7$  Instituto de F\'{i}sica Te\'orica (IFT-UAM/CSIC), Universidad Aut\'onoma de Madrid, 28049 Madrid, Spain\\
$^8$ Instituci\'o Catalana de Recerca i Estudis Avan\c{c}ats (ICREA), 08010 Barcelona, Spain\\
$^9$  CIEMAT, Centro de Investigaciones Energ\'{e}ticas, Medioambientales y Tecnol\'{o}gicas, Avda. Complutense 40, 28040 Madrid, Spain
}

\date{Accepted XXX. Received YYY; in original form ZZZ}

\pubyear{2015}

\begin{document}
\label{firstpage}
\pagerange{\pageref{firstpage}--\pageref{lastpage}}
\maketitle

\begin{abstract}
In this work, we explore the application of intensity mapping to detect extended Ly$\alpha$ emission from the IGM via cross-correlation of PAUS images with Ly$\alpha$ forest data from eBOSS and DESI. Seven narrow-band (FWHM=13nm) PAUS filters have been considered, ranging from 455 to 515 nm in steps of 10 nm, which allows the observation of Ly$\alpha$ emission in a range $2.7<z<3.3$. The cross-correlation is simulated first in an area of 100 deg$^2$ (PAUS projected coverage), and second in two hypothetical scenarios: a deeper PAUS (complete up to $i_{\rm AB}<24$ instead of $i_{\rm AB}<23$, observation time x6), and an extended PAUS coverage of 225 deg$^2$ (observation time x2.25). A hydrodynamic simulation of size 400 Mpc/h is used to simulate both extended Ly$\alpha$ emission and absorption, while the foregrounds in PAUS images have been simulated using a lightcone mock catalogue.
Using an optimistic estimation of uncorrelated PAUS noise, the total probability of a non-spurious detection is estimated to be 1.8\% and 4.5\% for PAUS-eBOSS and PAUS-DESI , from a run of 1000 simulated cross-correlations with different realisations of instrumental noise and quasar positions. The hypothetical PAUS scenarios increase this probability to 15.3\% (deeper PAUS) and 9.0\% (extended PAUS). With realistic correlated noise directly measured from PAUS images, these probabilities become negligible. Despite these negative results, some evidences suggest that this methodology may be more suitable to broad-band surveys.
\end{abstract}




\section{Introduction}\label{sec:Introduction}

In the last few years, the amount of observational data for the Universe at different wavelengths has steadily increased, which has led to the development of new methods and techniques to analyse these observations. Intensity mapping (IM) is one of these techniques, consisting of the tracing of large-scale structure with one or more emission lines, without resolving any kind of finite source, like galaxies or quasars. The use of a sharp and narrow spectral feature, such as an emission line, allows us to map the structure not only in angular coordinates but also in redshift, which provides 3-D tomography of the tracer \citep{Peterson2009}.

Originally, this technique was proposed to study the power spectrum with the 21-cm emission line at high, pre-reionization redshifts ($z>5$) \citep{Madau1997, Loeb2004}, but its application at lower redshifts has also been studied, e.g., as a method to measure Baryonic Acoustic Oscillations (BAO)  \citep{Chang2008}. Other emission lines have also been considered, such as the CO rotational line at intermediate \citep{Breysse2014, Li2016} or high redshift, \citep{Carilli2011},  CII emission line \citep{Gong2012, Yue2015}, or the Ly$\alpha$ line \citep{Silva2013, Pullen2014}. Given the short wavelength of this last line (121.567 nm), Ly$\alpha$ emission can only be observed at $z>2$ with ground-based telescopes, which limits any IM study with this tracer to relatively high redshifts.

Since IM does not resolve individual objects but considers all emission at certain wavelengths, one of the main challenges that IM studies face is contamination by foregrounds. This source of noise can be removed via cross-correlation with other datasets of objects with well-known redshift, an approach that has been successfully applied in detections of the 21-cm line \citep{Chang2010}, CII emission line \citep{Pullen2018} and the Ly$\alpha$ line \citep{Croft2016, Chiang2019}, coming from HI in the intergalactic medium (IGM).

In the Ly$\alpha$ case, in \citet{Chiang2019} detection of extragalactic background light (EBL) is reported using cross-correlation of UV broad-band data from the Galaxy Evolution Explorer (GALEX) with spectroscopic galaxy samples; with an adjustable spectral model of the EBL they place constraints on total Ly$\alpha$ emission up to $z=1$. This work only considers the evolution of EBL and its properties in redshift direction by integrating the cross-correlation in an angular range corresponding to 0.5-5 Mpc/h, so other than the redshift evolution, the results of this work are confined to cluster scales.

Regarding \citet{Croft2016} all data used for IM was extracted from the Sloan Digital Sky Survey III (SDSS-III, \citet{Eisenstein2011}) Baryon Oscillation Spectroscopic Survey (BOSS, \citet{Dawson2013}). Ly$\alpha$ emission is estimated by selecting spectra of Luminous Red Galaxies (LRGs)  at $z < 0.8$ and subtracting a best fit model for each galaxy spectrum, which leaves a significant amount of Ly$\alpha$ surface brightness from higher redshifts. These residual spectra are cross-correlated with quasars from the same catalogue, which gives a detection at mean redshift $z=2.5$ of large-scale structure at a $8\sigma$ level, and a shape consistent with the $\Lambda$CDM model. This cross correlation, however, only yields a positive signal on scales up to 15 Mpc/h. Given the quasar density of BOSS, this implies that only 3\% of the space (15 Mpc/h around quasars) is being mapped, and large scale structure of Ly$\alpha$ emission in general is not being constrained by this measurement. Ly$\alpha$ emission is  extended at high enough redshift (approximately $z>3$), with Ly$\alpha$ blobs \citep{Taniguchi2001, Matsuda2004} forming visible structures around quasars up to hundreds of kpc in size, and the integrated faint Ly$\alpha$ emission in turn covers almost 100\% of the sky \citep{Wisotzki2018}. Therefore, cross-correlation of the Ly$\alpha$ emission with a more suitable dataset (less rare than quasars) is expected to provide a positive signal on larger scales.

One of these possible datasets to cross-correlate with is the Ly$\alpha$ forest i.e., the set of absorption lines that appears in the spectrum of quasars due to the HI mass distribution between the object and the observer \citep{Rauch1998}. Each Ly$\alpha$ forest spectrum contains information about the HI distribution along a large fraction of the entire line of sight, which should allow cross-correlation over larger, more representative volumes. In \citet{Croft2018} a first attempt at  cross-correlation was performed between Ly$\alpha$ forest from BOSS and similar LRG spectra with the best galaxy fit subtracted to those used in \citet{Croft2016}, but no signal was found. Nonetheless, BOSS was not designed with Ly$\alpha$ IM as an objective, and it is certain that larger and more suitable datasets are needed to obtain a clear detection \citep{Kovetz2017}. Such a dataset would need data with redshift precision close to that achieved by spectroscopy over large areas, providing a volume large enough to study large scale structure with Ly$\alpha$ IM \citep{Croft2018}. One potential candidate that may fulfil these requirements are narrow-band imaging surveys, such as the Physics of the Accelerating Universe Survey (PAUS, \citealt{Castander2012,  Eriksen2019}).

The object of this work is to simulate the cross-correlation of PAUS images with Ly$\alpha$ forest data from two different spectroscopic surveys, in order to compute the two-point correlation function (2PCF), as well as to evaluate if meaningful constraints can be obtained. The spectroscopic surveys considered for this purpose are the already available SDSS extended Baryon Oscillation Spectroscopic Survey (eBOSS, \citet{Dawson2015}), and the upcoming Dark Energy Spectroscopic Instrument (DESI) Experiment \citet{DESICollaboration2016}.

For all the calculations in this paper the following flat cosmology has been assumed: $h=0.702$, $\Omega_{\rm m}=0.275$, $\Omega_\Lambda=0.725$, $\Omega_{\rm b}=0.046$, $n_s=0.968$, $\sigma_8=0.82$. This is the cosmology of the hydrodynamic simulation we have used to model the Ly$\alpha$ extended emission and the Ly$\alpha$ forest \citep{Ozbek2016}, which has also been used for the entirety of the work for the sake of consistency.

The paper is structured as follows. In Section \ref{sec:Surveys to cross-correlate} the two datasets to be cross-correlated (PAUS and eBOSS/DESI) are briefly summarised. Section \ref{sec:Simulation of the survey data} shows how these datasets are simulated by combining the aforementioned hydrodynamic simulation and a lightcone mock catalogue. In Section \ref{sec:Simulated cross-correlation estimator}, the estimator to compute the observed cross-correlation from the two datasets is explained, as well as some caveats to be taken into account for this particular case. Section \ref{sec:Theoretical correlation function} describes the theoretical calculation of the two-point correlation function from the matter power spectrum. Section \ref{sec:Results} shows the results from both the theoretical correlation function and the simulated cross-correlation; the bias of the extended Ly$\alpha$ emission/Ly$\alpha$ forest is derived from its comparison, and the likelihood of a cross-correlation detection is evaluated for different cases.  Finally, we conclude with Section \ref{sec:Conclusions}.

\section{Surveys to cross-correlate}\label{sec:Surveys to cross-correlate}

\subsection{PAUS}

\begin{figure}
 	\includegraphics[width=\columnwidth]{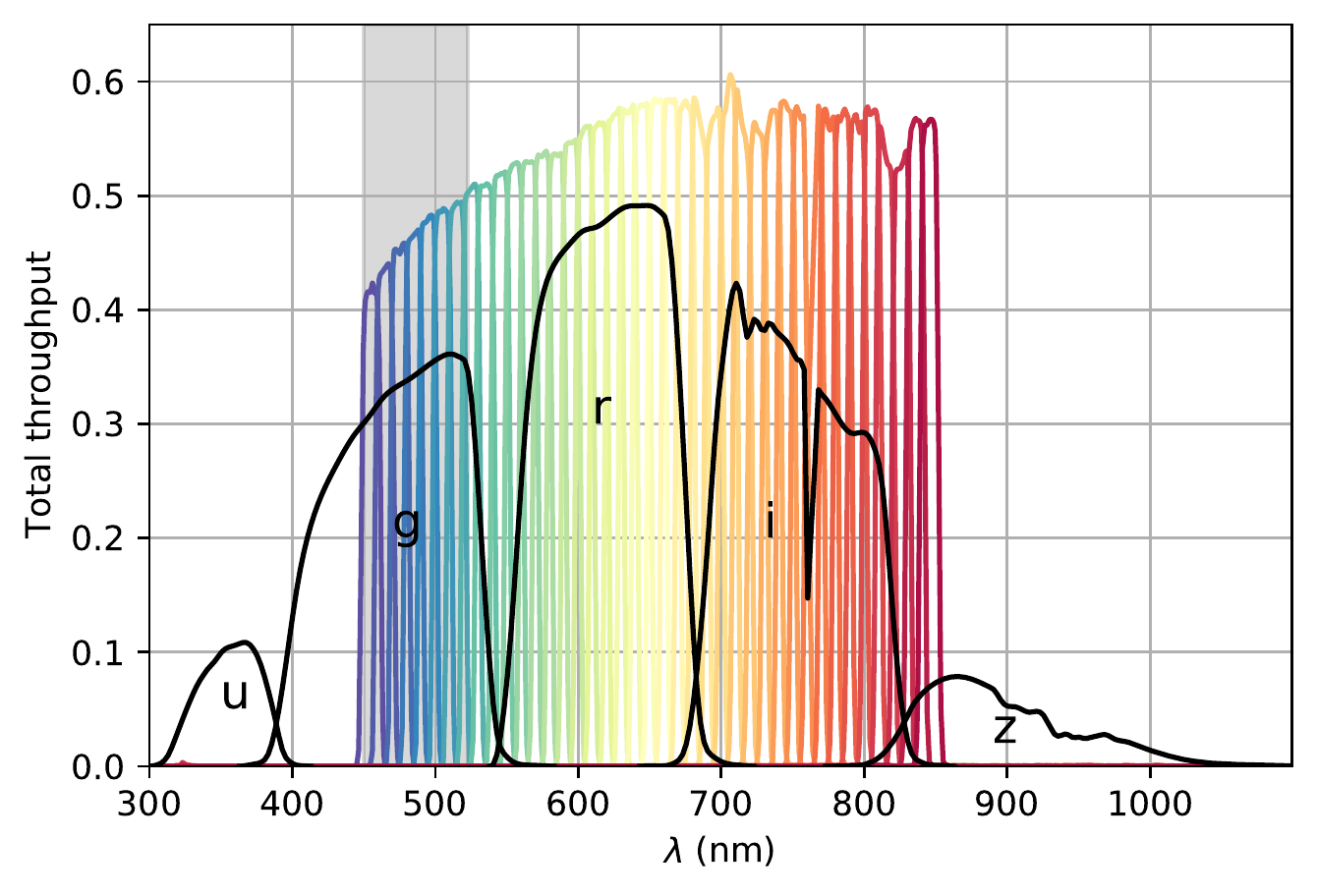}
     \caption{Response function for PAUS filters (coloured) and original SDSS $ugriz$ filters (black). Shaded area represents the wavelength range studied in this work.}
     \label{fig:PAUS_ugriz_filters}
\end{figure}

PAUS is a photometric imaging survey currently being carried out at the William Herschel Telescope with the PAU Camera \citep{Castander2012}, whose main feature is the use of 40 narrow-band filters with a full width at half maximum (FWHM) of $\simeq 13$ nm, with mean wavelengths of 455 to 845 nm in steps of 10 nm (Fig. \ref{fig:PAUS_ugriz_filters}). Such a configuration allows one to obtain photometric redshifts (photo-z) with sub-percent precision over large sky areas \citep{Marti2014}. Preliminary results \citep{Eriksen2019} already achieve better photo-z precision than state-of-the-art photo-z measurements in the COSMOS field \citep{Laigle2016}. 

Although the main purpose of the survey is the elaboration of high-density galaxy catalogues with high-precision redshifts for cross-correlations of lensing and redshift distortion probes \citep{Gaztanaga2012}, the narrow-band data from PAUS may also be used for intensity mapping. The background of PAUS images, where no objects are resolved, also contains valuable cosmological information. Given the wavelength range of the NB filters, Ly$\alpha$ luminosity is observed in the range $2.7 < z < 6$, distributed in 40 redshift bins, one per each NB filter. At this redshift range faint Ly$\alpha$ emission surrounds most objects \citep{Wisotzki2018}, but foreground contamination must be removed first in order to study it.

For this work, however, only the seven bluest NBs will be considered, which span from 455 to 515 nm (shaded in Fig. \ref{fig:PAUS_ugriz_filters}). With these seven blue filters, Ly$\alpha$ emission is observed over the range $2.7<z<3.3$, approximately. At higher filter wavelengths, the observed Ly$\alpha$ emission increases in redshift, thus being farther away and fainter. In addition to this, the fraction of quasars observed at $z>3.3$ is extremely small (Fig. \ref{fig:quasar_redshift_distribution}), which means that the amount of Ly$\alpha$ forest data sampling this space is also very limited. Therefore, adding extra filters only provides a volume for the cross-correlation with lower SNR in PAUS images, and scarcely sampled by the Ly$\alpha$ forest.

The fields targeted by the survey are, in addition to COSMOS, the W1, W2 and W3 fields from the Canada-France-Hawaii Telescope Lensing Survey (CFHTLenS, \citet{Heymans2012}). The sum of the angular area of all these fields is $\sim$130 deg$^2$, but since a full coverage of the CFHTLenS fields is not expected, a total angular area of $\sim$100 deg$^2$ of PAUS images will be considered for this work. 

\subsection{eBOSS/DESI}

Both eBOSS and DESI are large spectroscopic surveys, with coverage of $\sim$10.000 deg$^2$ in the first case, and $\sim$14.000 deg$^2$ in the latter. Since eBOSS already fully overlaps with the fields observed by PAUS, and DESI is planned to contain the entirety of eBOSS fields, the limit on the angular area sampled by the cross-correlation is determined solely by how much PAUS observes.

Similarly, the limit on the redshift precision of the Ly$\alpha$ line is also set by PAUS narrow-band filters, not the eBOSS and DESI spectrographs. For eBOSS and DESI, the lowest resolution $R=\lambda/\Delta \lambda$ is approximately 2000, while for PAUS  the maximum resolution achievable would be around 65 if its photometric data was to be compared against spectroscopy. Therefore, detailed modelling  of the spectral resolution of Ly$\alpha$ forest data is not required, since the redshift resolution of the cross-correlation will be limited by the PAUS images. As long as the simulated Ly$\alpha$ forest has higher redshift resolution than PAUS images by at least one order of magnitude, it is safe to assume that any change to the spectral resolution of the Ly$\alpha$ forest will not impact the results.

\begin{figure}
 	\includegraphics[width=\columnwidth]{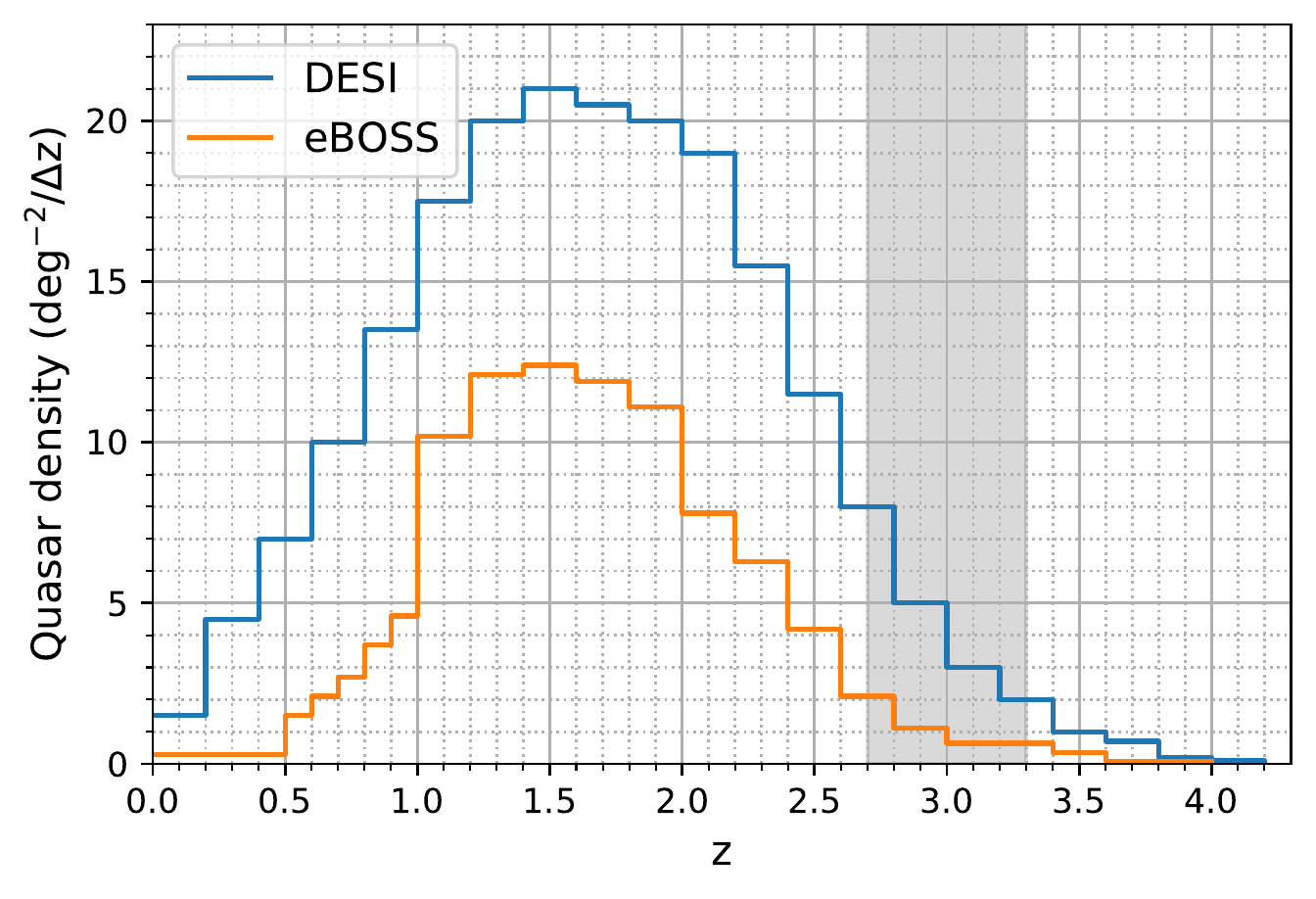}
     \caption{Projected quasar density vs redshift for DESI and eBOSS; shaded area shows the redshift interval of this study. The change of redshift binning at $z=0.1$ in eBOSS (from $\Delta z=0.1$ to $\Delta z=0.2$) is not relevant for this work.}
     \label{fig:quasar_redshift_distribution}
\end{figure}

Consequently, in order to know how much space is being sampled with the Ly$\alpha$ forest to simulate the cross-correlation, the only data needed from these surveys is the quasar density distribution with redshift. Fig. \ref{fig:quasar_redshift_distribution} shows this distribution for eBOSS and DESI, obtained respectively from \cite{Dawson2016}, Table 1 and \citet{DESICollaboration2016}, Fig. 3.17.

\section{Simulation of the survey data}\label{sec:Simulation of the survey data}

In order to simulate the cross-correlation between different surveys, the first step is to simulate the actual survey datasets. For this work, an already existing hydrodynamic simulation has been used for both Ly$\alpha$ forest data and Ly$\alpha$ emission, while the foregrounds in PAUS images have been computed using a broad-band mock catalogue interpolating the spectral energy distributions (SEDs) of objects by fitting SED templates. On top of the foregrounds, noise from any other sources (electronic, atmospheric, etc.) also needs to be modeled; this is done by assuming that the sum of all noise follows a Gaussian distribution, and measuring the variance of this distribution directly from PAUS reduced images.

This section is divided in two subsections. In Section \ref{sec:independent_simulations}, the three elements used for the modelled survey (hydrodynamic simulation, mock catalogue and noise) are described, and in Section \ref{sec:simulation_paus_lyalpha_im}, we explain how these datasets are combined to simulate both PAUS images and eBOSS/DESI Ly$\alpha$ forest data.

\subsection{Independent simulations}\label{sec:independent_simulations}

\subsubsection{Hydrodynamic simulation}\label{sec:hydrodynamic_simulation}

The hydrodynamic simulation used in this work has been performed with the P-GADGET code (\citealt{Springel2005, DiMatteo2012}), with $2 \cdot 4096^2$ particles in  a 400 Mpc/h box using the cosmology specified in \S \ref{sec:Introduction}. Particle masses of $1.19\cdot 10^7\, h^{-1} M_{\rm \odot}$ and $5.92 \cdot 10^7 h^{-1} M_{\rm \odot}$ were used for gas and dark matter respectively,  with a gravitational force resolution of $3.25 h^{-1} $ kpc. In order to speed up the simulation, the density threshold for star formation was lower than usual, so gas particles became collisionless star particles more quickly. This density threshold was 1000 times the mean gas density. Besides this, black hole formation and stellar feedback were not taken into account. While this results in inaccurate stellar properties of galaxies, it does not significantly affect the IGM, and thus the simulated Ly$\alpha$ forest \citep{Viel2004}.

\begin{figure*}
 	\includegraphics[width=\columnwidth]{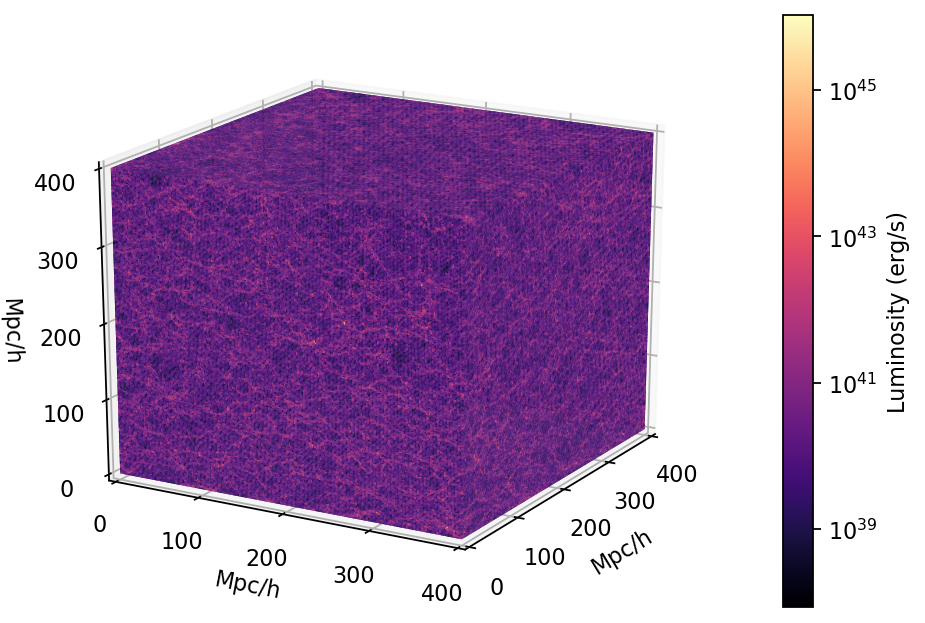}
 	\includegraphics[width=\columnwidth]{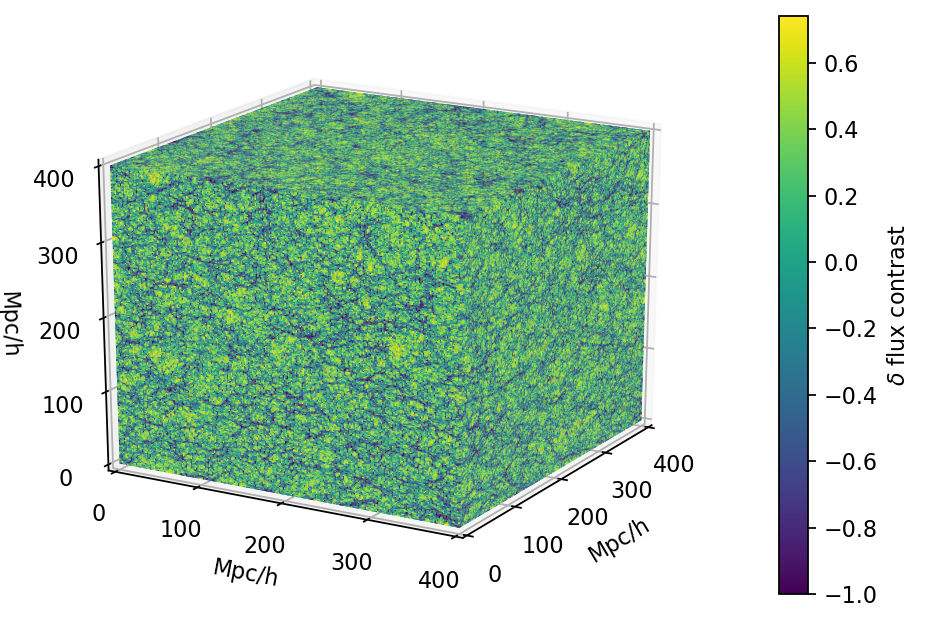}
     \caption{Hydrodynamic simulation used for this work. \textit{Left:} Extended Ly$\alpha$ emission, in erg/s and logarithmic colour scale. \textit{Right:} $\delta$ flux contrast, used to model the Ly$\alpha$ forest.}
     \label{fig:hydrodynamic_simulations}
\end{figure*}

This simulation was originally computed for Ly$\alpha$ forest studies, and has already been used in several works. In \citet{Cisewski2014} and \citet{Ozbek2016}, different methodologies to model the 3D IGM between Ly$\alpha$ forest data were tested with it, while in \citet{Croft2018} it was used to simulate Ly$\alpha$ IM. Fig. \ref{fig:hydrodynamic_simulations} shows a voxel plot of the hydrodynamic simulation in both Ly$\alpha$ emission, in luminosity units (erg/s), and absorption, in $\delta$ flux contrast, defined as

\begin{equation}
\delta_{\rm i} \equiv \frac{e^{-\tau_{\rm i}}}{\langle e^{-\tau} \rangle}-1.
\label{eq:delta_contrast}
\end{equation}
Where $\tau_{\rm i}$ is the optical depth of the Ly$\alpha$ forest pixel $i$, computed along sightlines through the simulation, as in \citet{hernquist96}. Therefore, with this definition high values of $\delta$ correspond to regions with low HI density, and vice-versa. This $\delta$ absorption flux is expected to have a clustering bias with respect to dark matter of $b_a=0.336\pm0.012$ at $z=2.25$ \citep{Slosar2011}, including redshift distortion effects.

While the physics leading to the Ly$\alpha$ forest absorption are reproduced explicitly in the hydrodynamic simulation, we make predictions for the Ly$\alpha$ luminosity using a simple heuristic model, with an amplitude
normalised using observational data; not enough is known about all sources of Ly$\alpha$ emission to warrant using a more detailed model.

In this model, the Ly$\alpha$ luminosity is proportional to the square of the baryonic density field at the scale of the spatial binning used for this work (1.56 Mpc/h). 
This is done with the following expression

\begin{equation}
L_{\rm Ly\alpha}(r)=C_L\rho_{\rm b}(r)^2,
\label{eq:lya_emission}
\end{equation}

where $\rho_{\rm b}(r)$ is the baryonic density field, and $C_L$ is a normalisation constant chosen in order to set the average Ly$\alpha$ luminosity density to $1.1\cdot10^{40}$erg/s/(Mpc)$^3$. This value of Ly$\alpha$ luminosity density is that measured from observed
Ly$\alpha$ emitters at redshift $z=3.1$ \citep{Gronwall2007}, which is a conservatively low
value to use, as it does not include any sources of
Ly$\alpha$ emission which are not readily observed in
narrow band Ly$\alpha$ surveys. This includes 
low surface brightness extended halos around Ly$\alpha$ 
emitters \citep[e.g.,][]{steidel11} (which could host 50\%
or more extra Ly$\alpha$ luminosity density), or
any other low surface brightness emission which could be difficult to detect in surveys aiming to detect objects
above a threshold, but which would be included in an intensity map.
The Ly$\alpha$ luminosity density we use can be converted to an associated star formation rate density applying a commonly used relation between Ly$\alpha$ luminosity and star formation rate (SFR) of $1.1 \cdot 10^{42}$ erg/s/($M_{\rm \odot}$/yr) at $z\sim3$ \citep{Cassata2011}. This relation yields a SFR density measured from observed Ly$\alpha$ emitters of 0.01 $M_{\rm \odot}$/yr/Mpc$^{3}$ \citep{Gronwall2007}.

Once the Ly$\alpha$ luminosity density is determined for a simulation cell in the model, we convolve the  Ly$\alpha$ luminosity values with the line of sight velocity field, in order to put the  Ly$\alpha$ emission into redshift
space. This technique is similar to that used to convert the Ly$\alpha$ forest absorption spectra
into redshift space \citep[see e.g.,][]{hernquist96}.

The baryons are unbiased with respect to dark matter, and thus in the model, the Ly$\alpha$ emission is expected to be biased with respect to dark matter by a factor $b_{\rm e} \sim 2$ on linear scales
(due to Ly$\alpha$ being related to the square of the baryonic density).
This $b_{\rm e}$ in the model is chosen to be consistent with the measured bias of Ly$\alpha$ emitters at these
redshifts \citep[e.g.,][]{Gawiser07},  considering that these are the predominant sources of Ly$\alpha$ emission, and that the contribution of the IGM is subdominant.
We note that the assumption of squaring the density will lead to a linear bias 
of $b_{\rm e}=2$ may not hold at very highest densities, and this may result in artefacts in the form of extremely bright pixels. As it is explained later (\S \ref{sec:paus_images_ly_alpha}), a Ly$\alpha$ flux threshold is set for the simulated PAUS images, partially in order to account for this effect.

\subsubsection{Mock catalogue/Foreground simulation}

If we consider PAUS images for Ly$\alpha$ IM, most of the detected photons of cosmic origin will not come from Ly$\alpha$ at a certain redshift (depending on the filter used), but from uncorrelated sources at different redshifts than the expected Ly$\alpha$ emission. The main contributors to this contamination of the signal will be foregrounds, i.e., objects with lower redshift between the Ly$\alpha$ emission and the observer. In this work, 96.7\% of all the observed flux in the simulation (averaged over all filters) was from foregrounds.

Since the objective of this paper is assessing the potential of cross-correlating PAUS with Ly$\alpha$ forest
data, a realistic model of these foregrounds is key for our study. In order to model them, we will need a mock catalogue that spans a range of redshift large enough (at least $z=2.75$, but ideally until $z=6$, where the PAUS redshift range for Ly$\alpha$ ends), with an angular size comparable to the Ly$\alpha$ forest/emission simulation box. Besides, all objects in the catalogue must have their observed SEDs in the PAUS wavelength range (455-855 nm) and with resolution higher than PAUS FWHM ($\Delta\lambda < 13$ nm).

The two first requirements (redshift range and angular size) are met by already available mock catalogues, but none of them contain direct SED information (at least, not to the best of the authors knowledge). Such mock catalogues are intended to reproduce large surveys, with the only spectral information available being either broad bands, which do not meet the resolution requirement, or emission lines, which are insufficient to generate the foregrounds.

Our approach to this problem is to take a mock catalogue with broad bands, and interpolate SEDs for all objects by fitting SED templates to the broad bands. The mock catalogue selected  is a lightcone originally developed to simulate data from the Euclid satellite, made from a run of the Millennium Simulation using WMAP7 cosmology \citep{Guo2013}. This lightcone is complete up to magnitude 27 in Euclid $H$ band, which makes it ideal for foreground simulations (since most mock catalogues do not reach such depths). The semi-analytical model applied to compute galaxies is \textsc{galform} \citep{Gonzalez-perez2014a}, and the lightcone was constructed with the technique described in \citet{Merson2013a}.

\begin{figure}
 	\includegraphics[width=\columnwidth]{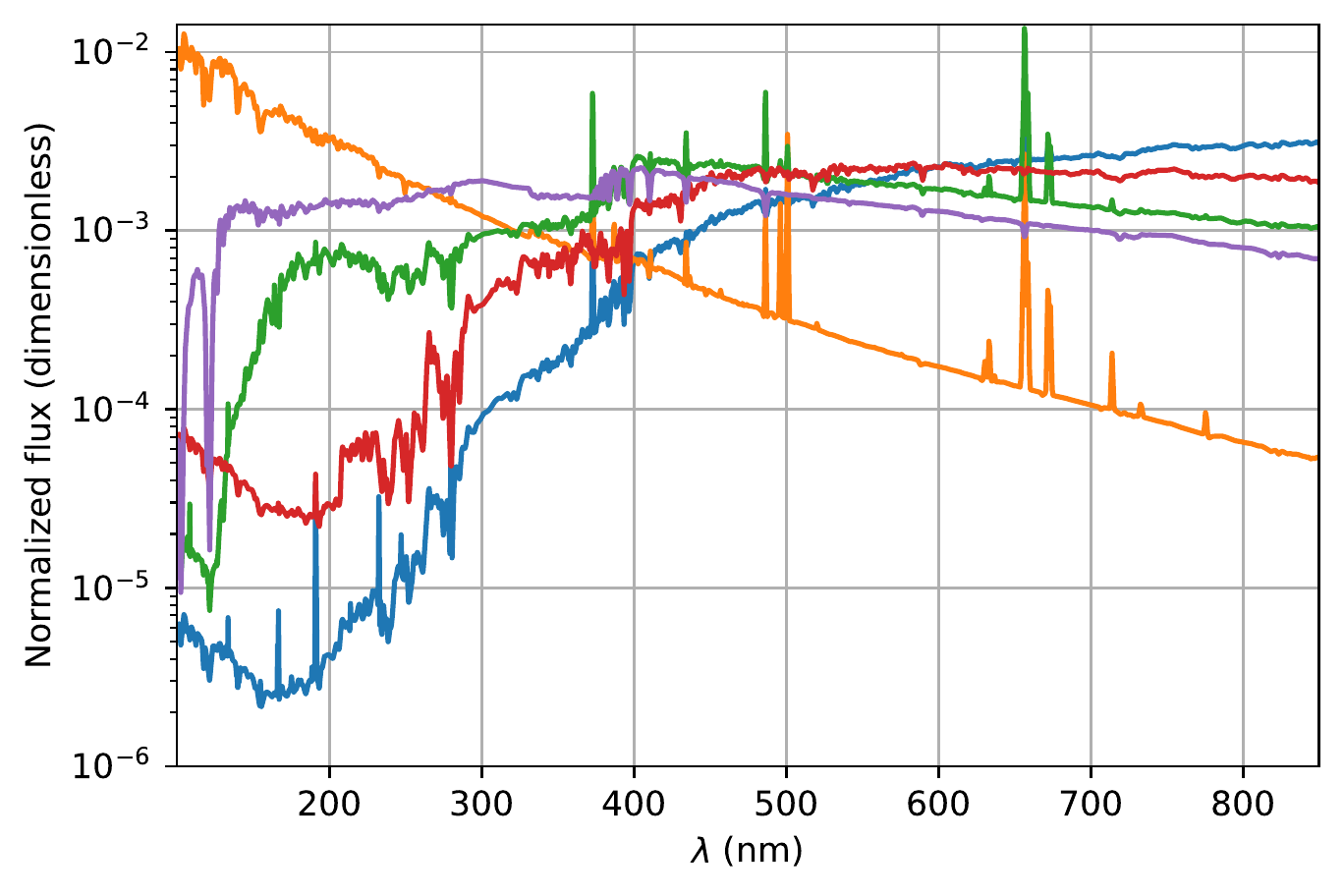}
     \caption{The five SED templates used for foreground simulation, normalised to facilitate visual comparison.}
     \label{fig:blanton_templates}
\end{figure}

In order to interpolate SEDs, we have considered the SDSS $ugriz$ bands (Fig. \ref{fig:PAUS_ugriz_filters}) from this mock and the five SED templates defined by \citet{Blanton2006}, which we show in Fig. \ref{fig:blanton_templates}. For the five templates, their $ugriz$ band values have been computed in a fine redshift grid ($\Delta z\sim0.01$). These template bands are used as the elements of a coordinate basis, and for any object the coefficients of the linear combination of templates that gives the $ugriz$ bands of the object can be computed with the following expression

\begin{equation}\label{eq:bands matrix}
    \left(
    \begin{matrix}
    u_{\rm obj} \\ g_{\rm obj} \\ r_{\rm obj} \\ i_{\rm obj} \\ z_{\rm obj}
    \end{matrix}
    \right)_z
    =
    \left(
    \begin{matrix}
    u_1 & g_1 & r_1 & i_1 & z_1 \\
    u_2 & g_2 & r_2 & i_2 & z_2 \\
    u_3 & g_3 & r_3 & i_3 & z_3 \\
    u_4 & g_4 & r_4 & i_4 & z_4 \\
    u_5 & g_5 & r_5 & i_5 & z_5 \\
    \end{matrix}
    \right)_{\rm z_{\rm grid}\sim z}
    \times
    \left(
    \begin{matrix}
    x_1 \\ x_2 \\ x_3 \\ x_4 \\ x_5
    \end{matrix}
    \right).
\end{equation}

Where the left hand array are the bands of the object $obj$ at redshift $z$, the right hand array $X$ are the coefficients of the linear combination of the templates, and the matrix is the basis of template bands at the redshift $z_{\rm grid}$ closest to the redshift of the object $z$. This is a simple linear system that has a single exact solution as long as the basis matrix is invertible (which has been checked for all $z_{\rm grid}$).
However, the coefficients $X$ must be all non-negative for the SED to make physical sense (since the SED templates are patterns of emitted flux for galaxies, and thus subtracting them has no physical meaning). Therefore, instead of finding the analytical solution, the coefficients are computed using non-negative least squares. This numerical method is approximate, but on average yields relative errors of a few percent when recovering the original bands. Once these coefficients are obtained, the linear combination of SED templates using the coefficients is computed for all objects, thus generating a full mock catalogue with high spectral resolution SEDs.

\subsubsection{PAUS Noise}

In addition to the foregrounds, PAUS images have noise from a large variety of sources (electronic, airglow, etc.), together with the intrinsic variability between nights (seeing, moonlight, etc). Instead of simulating each one of these components with a physical model, we have measured them directly from PAUS images. For each one of the 7 filters considered, 8 exposures in 10 different pointings in the COSMOS field have been blindly selected as a representative sample to evaluate the noise. All of these images were already reduced by the PAUS pipeline, but some additional processing was carried out to emulate the additional reduction that would be necessary for IM applications. 

First, resolved sources were removed by applying a sigma-clipping filter with 3$\sigma$ threshold in 5 iterations; ideally, the masks could be extracted from a reference catalogue, but as a preliminary study sigma-clipping is enough to virtually remove all resolved objects. The masked pixels were replaced by random values drawn from a Gaussian distribution with the same mean and $\sigma$ as the unmasked pixels of the image, to avoid having empty pixels that would result in an overestimation of $\sigma$ (since $\sigma$ needs  to be computed for the pixel size of the simulation, which is larger than the CCD pixel size, masked pixels would result in artificially smaller samples inside a simulation pixel, and thus a higher $\sigma$).

Second, once resolved objects were masked, the median flux value was computed and subtracted for all the images. This was done to cancel out the variability in sky brightness due to moon phases and time of observation, which may modify the average background flux by a factor of few. While this erases all Ly$\alpha$ clustering signal at scales larger than the CCD ($\sim$12 Mpc/h in its smallest dimension at $z=3$), this approach is enough for noise determination in this preliminary work. A proper modelisation of the moonlight and sky brightness as a function of date and time could remove this variability without erasing the large-scale Ly$\alpha$ signal, but it is out of the scope of this paper. However, it is a pending task if Ly$\alpha$ IM is to be performed on PAUS data (or other optical imaging surveys).

After this processing, $\sigma$ could be measured directly from the resulting images, but it would include not only the electronic and atmospheric noise aforementioned, but also the variance due to the cosmic foregrounds and the Ly$\alpha$ signal, which are already considered in our simulation. 

\begin{figure}
 	\includegraphics[width=\columnwidth]{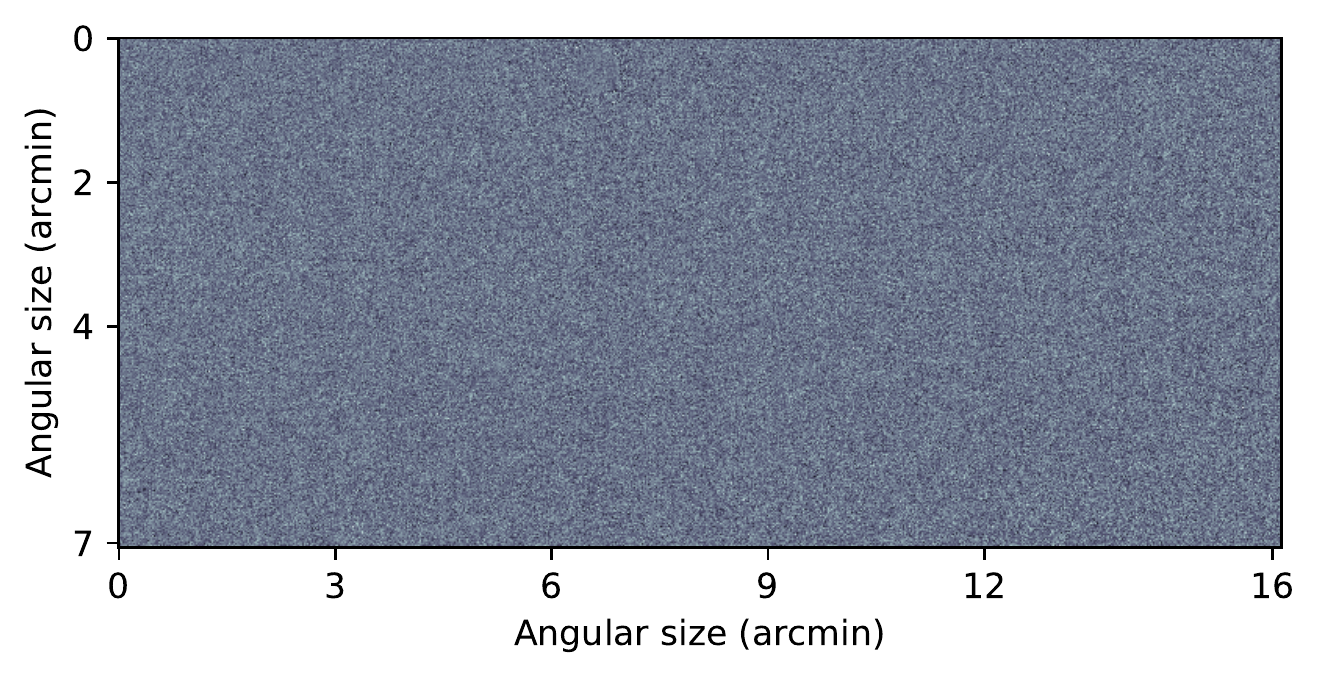}
     \caption{Example of stacked image in the 455 nm filter used to measure noise. 8 exposures of the same pointing have been stacked after subtracting the mean and applying sigma-clipping with 3$\sigma$.}
     \label{fig:stacked_image}
\end{figure}

In order to remove all signals from cosmic origin and keep only the electronic and atmospheric noise, we have stacked all exposures for all pointings, but applying an scaling factor of -1 to half of them. Since the number of exposures is even, any signal that should remain constant between exposures would tend to zero (e.g., both Ly$\alpha$ and foreground emission), while the variability due to atmospheric and electronic components remains. This has been done using \textsc{swarp} \citep{Bertin2002}, disabling background removal (to not artificially decrease the resulting noise), and cropping the regions of the stacked image where there was not a full overlap of all the 8 exposures. Fig. \ref{fig:stacked_image} show an example of the resulting stacked image.

For each one of these stacked images, the $\sigma$ was computing for increasing pixel sizes, starting by the intrinsic pixel size of the CCD, and going above the pixel size of our hydrodynamic simulation. The flux values of these increasing pixel sizes were computed by adding the values of all pixels inside them instead of averaging, since the hydrodynamic simulation considers the total Ly$\alpha$ luminosity in each 3D pixel, not its spatial average. Given that the images are stacked, the resulting $\sigma$ has been divided by $\sqrt{N_{\rm exp}}$, to scale the result to a single exposure. 

\begin{figure}
 	\includegraphics[width=\columnwidth]{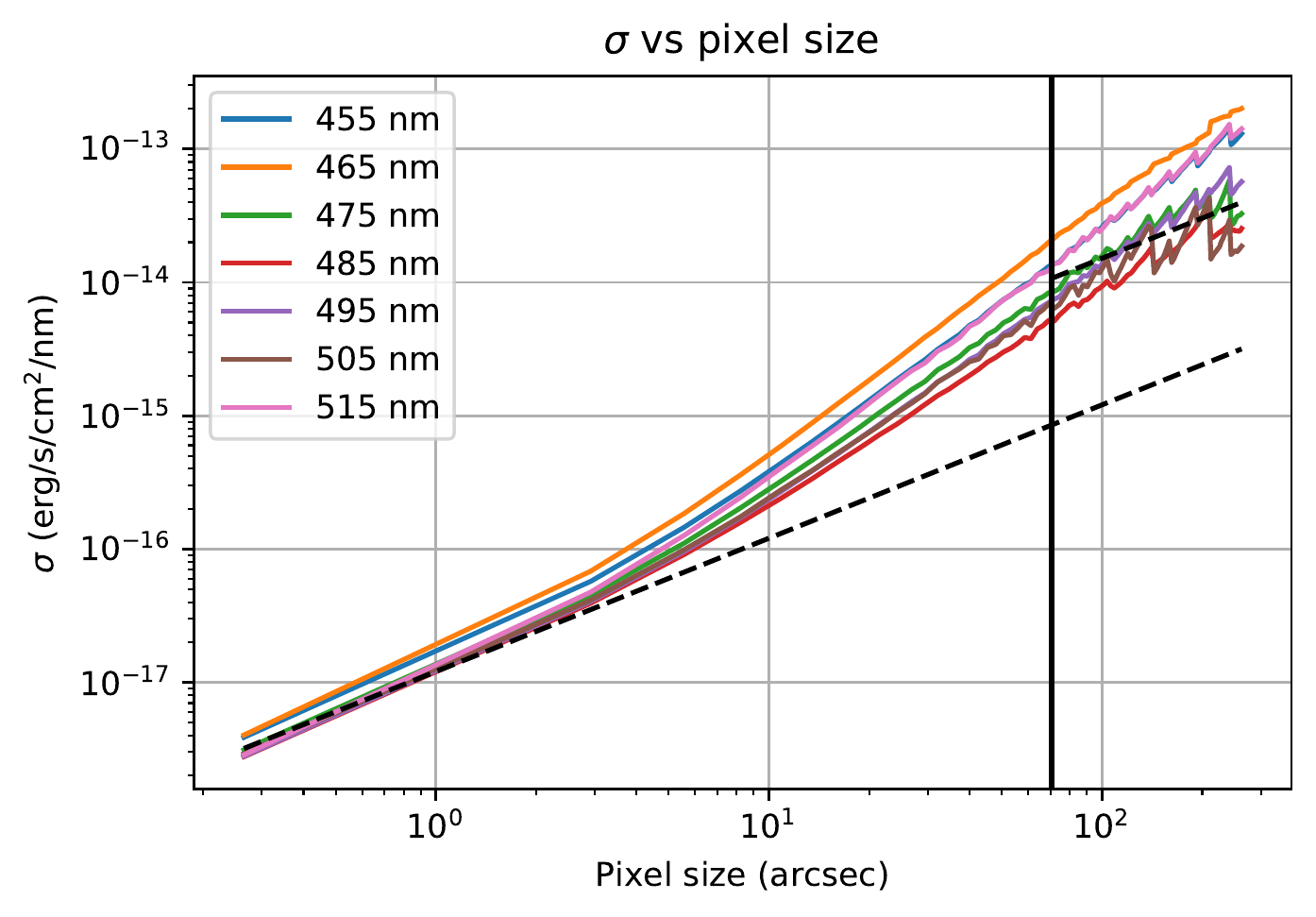}
     \caption{Average measured $\sigma$ of masked images vs pixel size, for the seven filters used in this work. The dashed line shows the extrapolated mean for uncorrelated noise, and the vertical line the pixel size of the simulation.}
     \label{fig:noise_sigma}
\end{figure}

Fig. \ref{fig:noise_sigma} shows the average $\sigma$ for each filter versus pixel size; the vertical line represents the pixel size of the simulation, and the dashed line an extrapolation of the $\sigma$ vs pixel size considering uncorrelated noise (averaged for the 7 filters). This extrapolation has been determined with 

\begin{equation}\label{eq:sigma_extrapolation}
\sigma_1 = \frac{\theta_1}{\theta_0} \sigma_0,
\end{equation}

where $\sigma_0$ and $\theta_0$ are the standard deviation and angular size for the original pixels of the image, and $\sigma_1$, $\theta_1$ its counterparts for the new pixels. This expression comes from the fact that the sum of uncorrelated Gaussian variables has a $\sigma^2$ equal to the sum of all the $\sigma^2$ of the individual Gaussian distributions. In this approximation, we consider each pixel to be an uncorrelated Gaussian with equal $\sigma_{\rm pixel}$; thus, the sum of $n$ pixels will have $\sigma_{\rm sum}=\sqrt{n}\sigma_{\rm pixel}$. Since we are adding CCD pixels to form larger pixels where $\sigma$ is computed, this $\sqrt{n}$ factor will be equal to the ratio of angular sizes, which results in Eq. \ref{eq:sigma_extrapolation}.

By looking at Fig. \ref{fig:noise_sigma}, it is clear that the $\sigma$ measured from the images displays a noticeable correlation for pixel scales larger than 30 arcsec, since it shows a much steeper slope in logarithmic scale. With the approach we have followed, it is certain that this correlation is not of cosmic origin, but other than that, we can not speculate more on the causes for this observed correlation, which are left for future research.

Therefore, we have considered two different noise levels: first, the measured $\sigma$ for the simulation pixel size, which would be the intersection of the coloured lines with vertical black line in Fig. \ref{fig:noise_sigma}, and the uncorrelated $\sigma$ following Eq.\ref{eq:sigma_extrapolation}. The former represents the most realistic case if the cross-correlation with actual data were to be computed now, while the latter is an hypothetical case where through further work on image reduction all noise correlations are removed, and only the uncorrelated and irreducible electronic noise remains.

In Table \ref{tab:sigma_noise}, the mean $\sigma_{\rm noise}$ for each filter, measured at the pixel size of the simulation, as well as the scaled noise for three exposures $\sigma_{\rm 3exp}$, is shown, by dividing $\sigma_{\rm noise}$ by a factor of $\sqrt{N_{\rm exp}}$. A hypothetical case for a deeper PAUS (complete up to $i_{\rm AB}<24$) is also considered, since it is a possibility currently being explored. This would imply multiplying by six the current exposure time for all survey pointings, hence  the $\sigma_{\rm 18exp}$. Table \ref{tab:sigma_noise_uncorrelated} shows the same data for the uncorrelated noise approximation. Overall, the correlation in the noise increases $\sigma_{\rm noise}$ by a factor of $\sim$10.

\begin{table}
	\centering
	\caption{$\sigma_{\rm noise}$ measured for the pixel size of the simulation, in (erg/s/cm$^2$/nm)$\cdot 10^{-15}$, for the seven narrow-band filters, as well as its value scaled for three exposures, $\sigma_{\rm 3exp}$, and 18 exposures, $\sigma_{\rm 18exp}$.}
 	\label{tab:sigma_noise}
 	\begin{tabular}{rccccccc}
 		\hline
 		$\lambda$ (nm) & 455 & 465 & 475 & 485 & 495 & 505 & 515\\
 		\hline
 		$\sigma_{\rm noise}$  & 13.52 & 20.78 & 8.46 & 5.15 & 7.29 & 6.62 & 13.14\\
        $\sigma_{\rm 3exp}$ & 7.80 & 12.00 & 4.89 & 2.97 & 4.21 & 3.82 & 7.58\\
        $\sigma_{\rm 18exp}$  & 3.19 & 4.90 & 1.99 & 1.21 & 1.72 & 1.56 & 3.10\\
        \hline
 	\end{tabular}
 \end{table}
 
 \begin{table}
	\centering
	\caption{$\sigma_{\rm noise}$ extrapolated as uncorrelated noise to the pixel size of the simulation, in (erg/s/cm$^2$/nm)$\cdot 10^{-16}$, for the seven narrow-band filters, as well as its value scaled for three exposures, $\sigma_{\rm 3exp}$, and 18 exposures, $\sigma_{\rm 18exp}$.}
 	\label{tab:sigma_noise_uncorrelated}
 	\begin{tabular}{rccccccc}
 		\hline
 		$\lambda$ (nm) & 455 & 465 & 475 & 485 & 495 & 505 & 515\\
 		\hline
 		$\sigma_{\rm noise}$  & 10.32 & 10.74 & 8.32 & 7.39 & 7.46 & 7.79 & 7.63\\
        $\sigma_{\rm 3exp}$ & 5.96 & 6.20 & 4.81 & 4.27 & 4.31 & 4.50 & 4.40\\
        $\sigma_{\rm 18exp}$  & 2.43 & 2.53 & 1.96 & 1.74 & 1.76 & 1.84 & 1.80\\
        \hline
 	\end{tabular}
 \end{table}

\subsection{Simulation of PAUS Ly$\alpha$ IM}\label{sec:simulation_paus_lyalpha_im}

\subsubsection{PAUS images: Ly$\alpha$ emission}\label{sec:paus_images_ly_alpha}
In order to simulate the PAUS images for the cross-correlation, the elements explained in the previous subsection (Ly$\alpha$ emission from the hydrodynamic simulation, foregrounds from the mock catalogue and Gaussian noise) must be converted to units of observed flux density (erg/s/cm$^2$/nm) and merged into the seven narrow-band filters.

Since the hydrodynamic simulation gives Ly$\alpha$ emission in luminosity units (erg/s), the first step is to compute the comoving coordinates of all pixels of the simulation from the point of view of the observer. Assuming the cosmology of the simulation, and knowing that the simulation snapshot is at $z=3$, we consider the comoving distance from the observer to the centre of the box to be the radial comoving distance at redshift 3, $\chi(z=3)$. Knowing this, the comoving coordinates of all cells of the simulation with respect to the observer are also known (as well as their edges), assuming that the three axes of the simulation box are RA, dec and radial directions respectively. The bins of the hydrodynamic simulation are not in spherical coordinates but Cartesian, however, given the small angular size of the sample, the small-angle approximation can be applied.

With the comoving radial distance of all cells known, and the relation $\chi(z)$ given by the cosmology, the inverse relation $z(\chi)$ can be computed numerically, and thus a redshift can be assigned to each cell. This allows to compute the luminosity distance simply with its definition for a flat cosmology

\begin{equation}\label{eq:luminosity-comoving_distance}
D_L(z) = (1+z)\cdot \chi(z).
\end{equation}

Moreover, given that all the emitted flux is Ly$\alpha$, the rest frame wavelength is also known ($\lambda_{\rm Ly\alpha}=121.567$ nm), which yields the observed wavelength range of all cells in the hydrodynamic simulation, and thus all redshift bins (following the small angle approximation, all cells in the same radial distance bin will have the same redshift, and thus observed wavelength range). With all these elements computed, the observed flux density for all PAUS cells comes from the following expression

\begin{equation}\label{eq:luminosity_to_observed_flux_density}
f_{\rm \lambda\, i}= \frac{L}{4\pi {D_L(z_{\rm i})}^2 \Delta\lambda_{\rm i}^{\rm obs}}.
\end{equation}

Where L is the cell luminosity given by the hydrodynamic simulation (erg/s), $D_L$ the luminosity distance in cm, and $\Delta\lambda_{\rm i}^{\rm obs}$ the observed wavelength range for the redshift bin of the cell, all corresponding to the PAUS cell $i$.

Having computed the observed flux density for all PAUS cells, the redshift bins of the PAUS simulation need to be merged to simulate the wavelength bins given by PAUS filters. In order to do so, PAUS filters are considered to have top-hat response functions 10 nm wide, ranging from 455 nm (bluest filter) to 845 nm (reddest). Following this criterion, the redshift bins of the simulation completely fill the seven bluest filters, which also limits the cross-correlation to seven filters in this work. The last four redshift bins of the simulation fall outside the seventh filter; these bins are discarded for the simulation of PAUS images. For each one of the simulated PAUS filters, all the redshift bins of the hydrodynamic simulation that fall inside the wavelength range of the filter are merged into a single one, with its flux value being the mean of the merged bins (since observed fluxes are the average flux density over the response function).

With redshift bins already merged to simulate PAUS filters, the average Ly$\alpha$ redshift for each filter can be used to convert from observed flux densities (erg/s/cm$^2$/nm) to absolute flux densities (erg/s/nm), with the following expression

\begin{equation}\label{eq:observed_to_absolute_flux}
F_{\rm \lambda \, i}= 4\pi {D_L(z_{\rm nb})}^2 f_{\rm \lambda\, i}.
\end{equation}

Where $z_{\rm nb}$ is the redshift of Ly$\alpha$ in the respective narrow band. This is done in order to cancel out the dimming of observed Ly$\alpha$ flux with redshift (due simply to the increasing distance between said emission and the observer), which would introduce an artificial gradient in the emission field to be cross-correlated. However, the previous conversion to observed fluxes was necessary, since we can only convert to absolute fluxes with observational data using the observed redshift, i.e., PAUS redshift bins, not the much finer redshift bins of the original simulation.

On top of this conversion to absolute fluxes, a realistic threshold can be imposed to Ly$\alpha$ fluxes, both to remove possible artefacts that may be derived from the assumption that Ly$\alpha$ luminosity is proportional to baryon density squared, and also to account for the fact that resolved objects will be removed from PAUS images before cross correlating (which may remove some bright Ly$\alpha$ emitters at high redshift).

\begin{figure}
 	\includegraphics[width=\columnwidth]{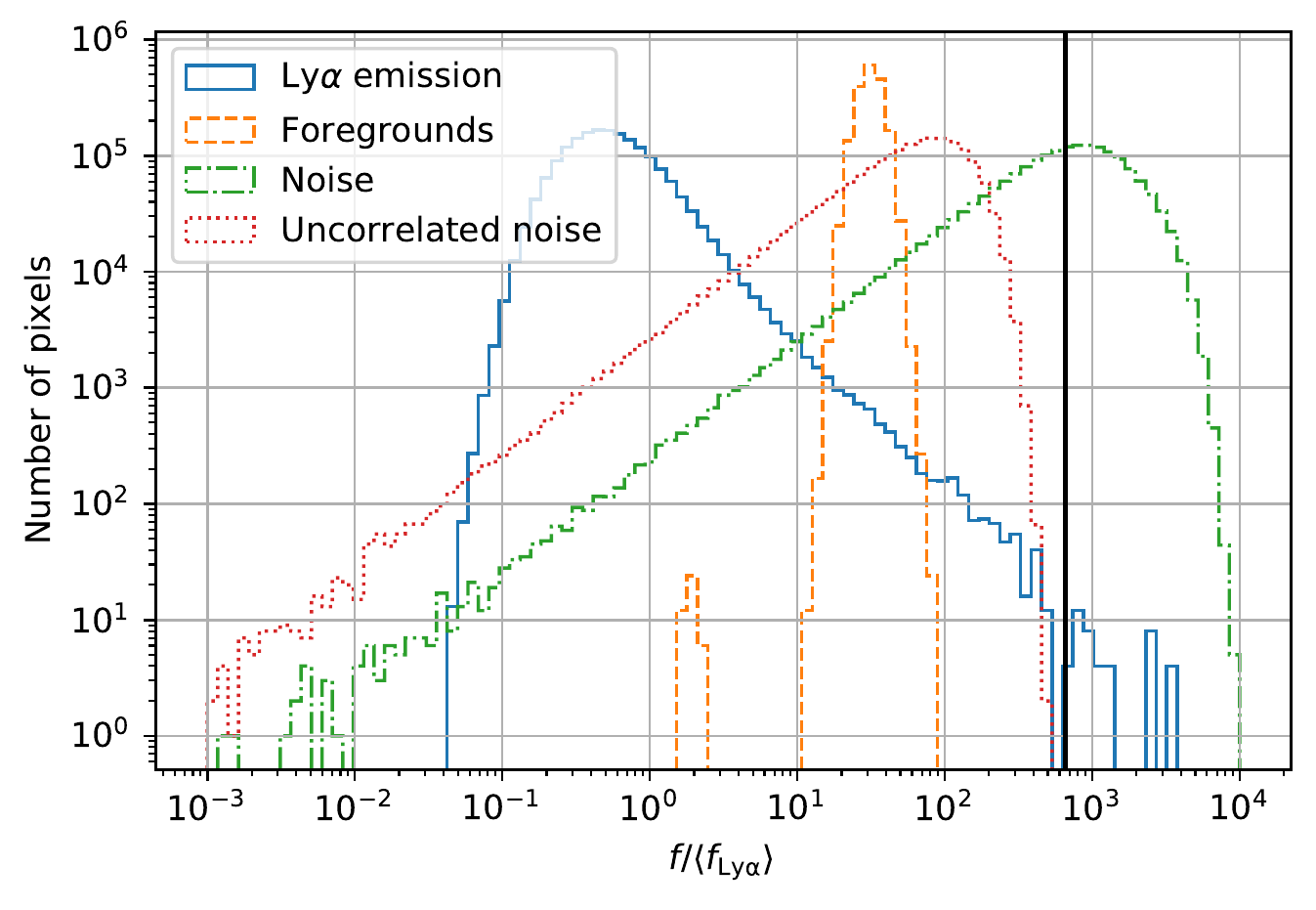}
     \caption{Logarithmic histograms of the ratio between fluxes and mean Ly$\alpha$ fluxes, for the Ly$\alpha$ emission, foreground emission, and instrumental noise, both the measured noise and the uncorrelated extrapolation (in absolute value). Noise values for the current PAUS case, $\sigma_{\rm 3exp\, abs}$. The vertical line represents the imposed Ly$\alpha$ threshold.}
     \label{fig:flux_histogram}
\end{figure}

The chosen Ly$\alpha$ absolute flux threshold is 10 times the brightest pixel of the simulated foregrounds, whose computation will be explained in \S \ref{sec:PAUS_images_foregrounds}. This value is chosen assuming that the foreground simulation gives a realistic estimate of how much unresolved flux can be expected, and taking into account that resolved objects are masked based on their $g$ band luminosity. This broad band has FWHM=138.7 nm (Fig. \ref{fig:PAUS_ugriz_filters}), which is one order of magnitude wider than PAUS narrow bands. Therefore, Ly$\alpha$ emission observed in a PAUS filter will be reduced by a factor of 10 when observed in the $g$ filter. A maximum value of of $1.53\cdot 10^{-5}$ erg/s/nm was set as a threshold, which affected only 0.0024 \% of all pixels. To visualise the extent of this threshold, Fig. \ref{fig:flux_histogram} shows histograms of absolute fluxes for the Ly$\alpha$ emission, foregrounds and instrumental noise, divided by the mean Ly$\alpha$ flux and together with the Ly$\alpha$ threshold, represented as a vertical line.

\begin{figure}
 	\includegraphics[width=\columnwidth]{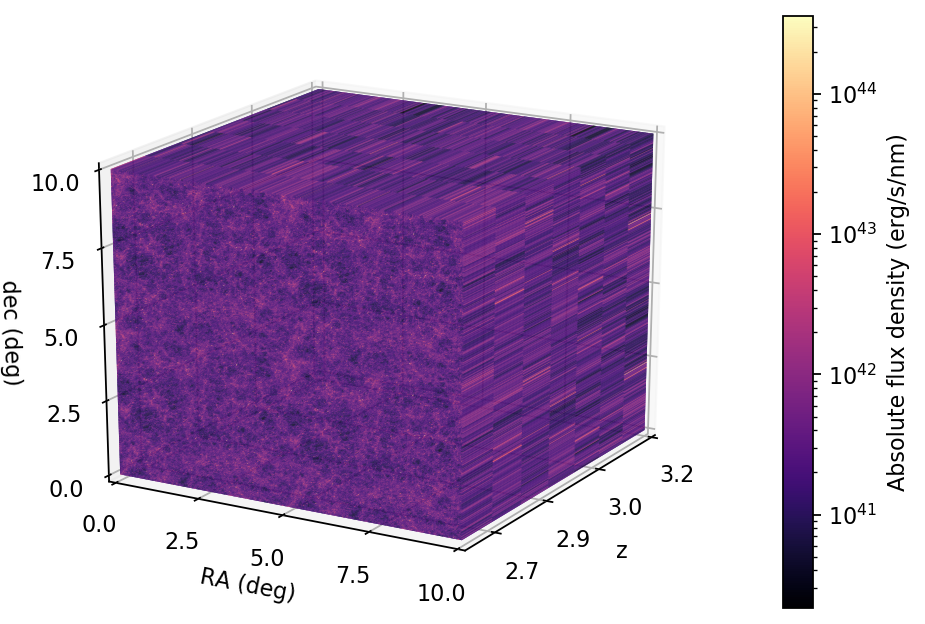}
 	\includegraphics[width=\columnwidth]{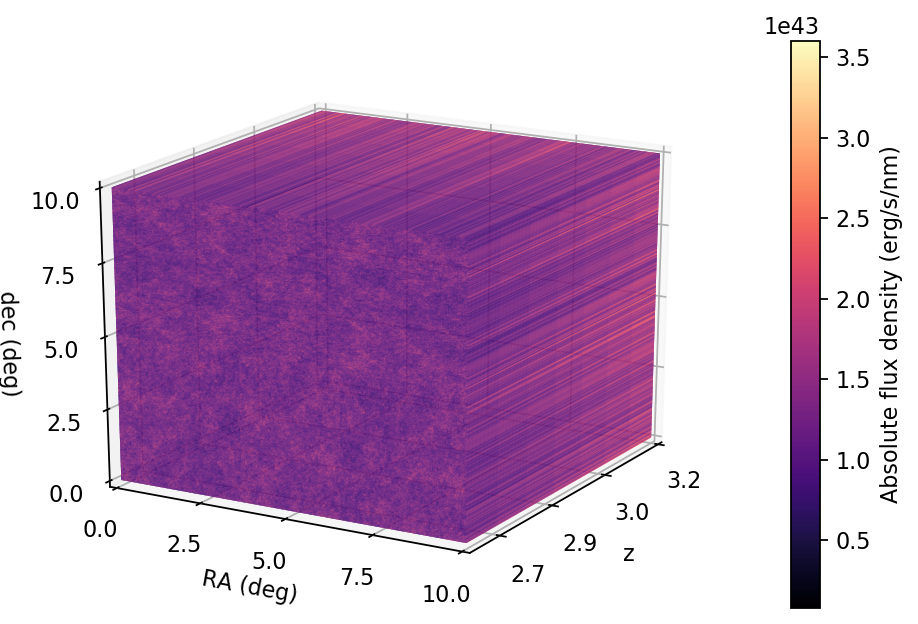}
     	\includegraphics[width=\columnwidth]{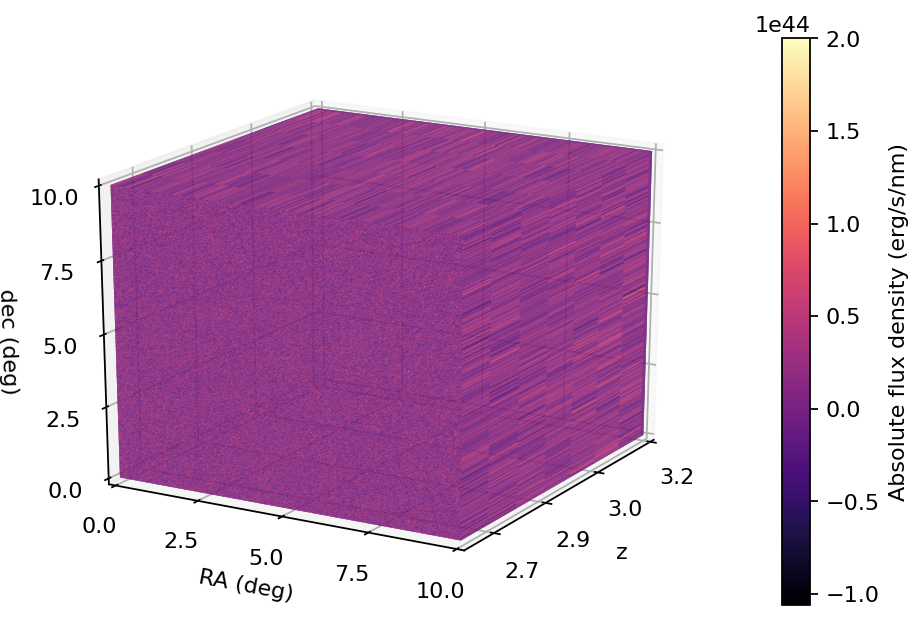}
     \caption{Simulation of PAUS images, in comoving coordinates according to Ly$\alpha$ observed redshift. Absolute flux densities in erg/s/nm. \textit{Top panel:} Ly$\alpha$ flux. \textit{Middle panel:} Foregrounds flux. \textit{Bottom panel:} Combined Ly$\alpha$ flux and foregrounds flux with instrumental noise (for the current PAUS case, $\sigma_{\rm 3exp\, abs}$, using the uncorrelated noise approximation).}
     \label{fig:PAUS_simulations}
\end{figure}

After all these steps, the result is a simulation of Ly$\alpha$ extended emission in PAUS filters. However, given the redshift and the size of the simulation, it only covers $\sim25$ deg$^2$, with an angular pixel size of 1.38 arcmin$^2$; since the expected area to cross-correlate is 100 deg$^2$, the simulation is replicated four times in mosaic pattern, which effectively covers the expected area. The result can be seen in Fig. \ref{fig:PAUS_simulations}, top panel.

\subsubsection{PAUS images: foregrounds}\label{sec:PAUS_images_foregrounds}

Given that resolved objects will be removed from PAUS images before cross-correlating, only the objects too faint to be resolved must be included in the foreground simulation. The PAUS reference catalogue is complete up to magnitude 25 in the $g$ band; consequently, only the objects in the mock lightcone dimmer than this value are selected. Besides this, since the lightcone is elliptical in angular coordinates, it is cropped to the largest inscribed rectangle. This rectangle is smaller than the $~25$ deg$^2$ at z$\sim$3 of the hydrodynamic simulation, so it is repeated in a mosaic pattern and cropped to cover the same angular area as the original Ly$\alpha$ simulation.
 
All the foreground objects have their SEDs computed by template fitting, as explained in the previous subsection, and they are binned in RA and dec using the same angular bins as the Ly$\alpha$ flux simulation. Since the templates are fitted to apparent magnitudes, by using the definition of AB magnitude the interpolated SEDs are already in observed flux units of erg/s/cm$^2$/nm.
 
For each one of these RA x dec pixels, the net observed SED is computed as the sum of the SEDs inside the bin. These stacked SEDs are then integrated and averaged over the response functions of the seven blue filters according to the expression below, which gives the observed foreground flux,
 
 \begin{equation}\label{eq:band_flux_integral}
f_{\rm nb}= \frac{\int_{\rm 0}^\infty d\lambda f_{\rm SED}(\lambda) R_{\rm nb}(\lambda)}{\int_{\rm 0}^{\infty}d\lambda R_{\rm nb}(\lambda)}.
\end{equation}

Here $f_{\rm SED}$ is the flux density of the interpolated SED, $R_{\rm nb}$ the response function of a certain narrow band, and $f_{\rm nb}$ the observed flux density in that narrow band. With this expression the observed foreground flux in the PAUS filters is obtained; in order to convert to absolute fluxes Eq. (\ref{eq:observed_to_absolute_flux}) is used.

The result is a three-dimensional array covering $\sim$25 deg$^2$ that can be directly added to the Ly$\alpha$ observed flux simulation. As in the Ly$\alpha$ flux case, this array needs to be replicated four times in a mosaic pattern for an effective coverage of 100 deg$^2$. This time, however, for each replication the array is rotated clockwise (keeping the redshift direction the same), in order to ensure that each 25 deg$^2$ subset is a different realisation of Ly$\alpha$ emission+foregrounds (if the rotation was not performed, replicating the arrays for a 100 deg$^2$ would be analogous to sampling the same 25 deg$^2$ area four times). The result of these simulated foregrounds can be seen in Fig. \ref{fig:PAUS_simulations}, middle panel.
 
This rotation introduces discontinuities in the foreground structure, since the periodic boundary conditions of the mock catalogue are broken. Nevertheless, the cross-correlation is computed by selecting cubes of PAUS cells around forest cells, so only forest cells close enough to the discontinuities will be affected by them. 

As shown in Fig. \ref{fig:smoothing_comparison}, the cross-correlation is only computed in perpendicular (angular) direction up to 20 Mpc/h. Given that the whole angular size of the simulation is ~800 Mpc/h, and that the discontinuities are two straight lines dividing the simulation in RA and dec, this leaves <10\% of the forest cells potentially affected by the discontinuities. Also, the dominant noise contribution is instrumental noise, not the foregrounds, so even in the small fraction of forest cells affected by discontinuities the effects of these on the cross-correlation should be fairly small.

\subsubsection{PAUS images: Combination and noise}

Considering that both have the same units and the same binning, the Ly$\alpha$ and foregrounds absolute flux simulations can be directly added into a total absolute flux array. The only step left to properly simulate PAUS observations is to add the instrumental and atmospheric noise. For this simulation, we have modeled this noise as Gaussian distribution of mean zero and $\sigma$ dependent on the filter. This $\sigma$ is the instrumental noise directly measured from images and scaled for the number of exposures, as specified in Table \ref{tab:sigma_noise}, converted to absolute flux units according to Eq. (\ref{eq:observed_to_absolute_flux}). Two cases have been considered: the $\sigma$ measured at the pixel size of the simulations (Table \ref{tab:sigma_noise}), and the extrapolation considering uncorrelated noise (Table \ref{tab:sigma_noise_uncorrelated}), which yields lower values of $\sigma$.

These absolute flux noise values, $\sigma_{\rm noise\, abs}$,  as well as the scaled value that is used,  $\sigma_{\rm 3exp\, abs}$, and the hypothetical deep PAUS, $\sigma_{\rm 18exp\, abs}$, are displayed in Table \ref{tab:sigma_noise_abs} for the real noise case, and in Table \ref{tab:sigma_noise_uncorrelated_abs} for the uncorrelated extrapolation. The final result of Ly$\alpha$ flux+foregrounds+instrumental noise is shown in Fig. \ref{fig:PAUS_simulations}, bottom panel. Only the uncorrelated noise case is shown, as the resulting figure in this case is already noise-dominated.

\begin{table}
\centering
	\caption{$\sigma_{\rm noise\, abs}$ with noise correlation, in (erg/s/nm)$\cdot10^{44}$, for the seven narrow-band filters, as well as its value scaled for three exposures, $\sigma_{\rm 3exp\, abs}$, and 18 exposures, $\sigma_{\rm 18exp\, abs}$.}
 	\label{tab:sigma_noise_abs}
 	\begin{tabular}{rccccccc}
 		\hline
 		$\lambda$ (nm) & 455 & 465 & 475 & 485 & 495 & 505 & 515\\
 		\hline
 		$\sigma_{\rm noise\, abs}$  & 9.12 & 15.05 & 6.56 & 4.27 & 6.45 & 6.24 & 13.16\\
        $\sigma_{\rm 3exp\, abs}$ & 5.27 & 8.69 & 3.79 & 2.47 & 3.72 & 3.60 & 7.60\\
        $\sigma_{\rm 18exp\, abs}$ & 2.15 & 3.55 & 1.55 & 1.01 & 1.52 & 1.47 & 3.10\\
		\hline
 	\end{tabular}
 \end{table}

\begin{table}
\centering
	\caption{$\sigma_{\rm noise\, abs}$ following the uncorrelated extrapolation, in (erg/s/nm)$\cdot10^{43}$, for the seven narrow-band filters, as well as its value scaled for three exposures, $\sigma_{\rm 3exp\, abs}$, and 18 exposures, $\sigma_{\rm 18exp\, abs}$.}
 	\label{tab:sigma_noise_uncorrelated_abs}
 	\begin{tabular}{rccccccc}
 		\hline
 		$\lambda$ (nm) & 455 & 465 & 475 & 485 & 495 & 505 & 515\\
 		\hline
 		$\sigma_{\rm noise\, abs}$  & 6.96 & 7.78 & 6.46 & 6.13 & 6.60 & 7.34 & 7.64\\
        $\sigma_{\rm 3exp\, abs}$ & 4.02 & 4.49 & 3.73 & 3.54 & 3.81 & 4.24 & 4.41\\
        $\sigma_{\rm 18exp\, abs}$ & 1.64 & 1.83 & 1.52 & 1.44 & 1.56 & 1.73 & 1.80\\
		\hline
 	\end{tabular}
 \end{table}

With this simulation, despite repeating both the Ly$\alpha$ emission and the foregrounds in a mosaic pattern, we ensure that the cross-correlation always samples a different combination of signal+noise, since instrumental noise is generated for the full simulation and foregrounds are rotated. 

While it may be argued that the clustering signal from Ly$\alpha$ emission is repeated, the only caveat of this is that cosmic variance may be underestimated. Given that the original diameter of the hydrodynamic simulation is 400 Mpc/h, far above the homogeneity scale \citep[e.g.,][]{Goncalves2018}, and that the predominant sources of noise are by far foregrounds and instrumental noise (as seen in \S \ref{sec:detection_probability_uncorrelated}), any effect cosmic variance may have on the result is negligible.

\subsection{eBOSS/DESI: Ly$\alpha$ forest}

\begin{figure*}
 	\includegraphics[width=\columnwidth]{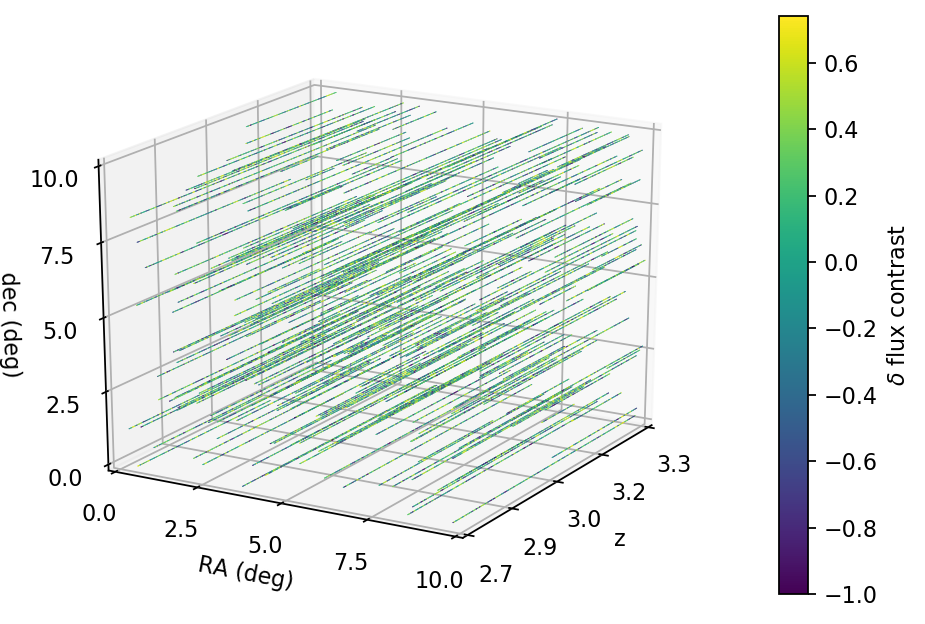}
 	\includegraphics[width=\columnwidth]{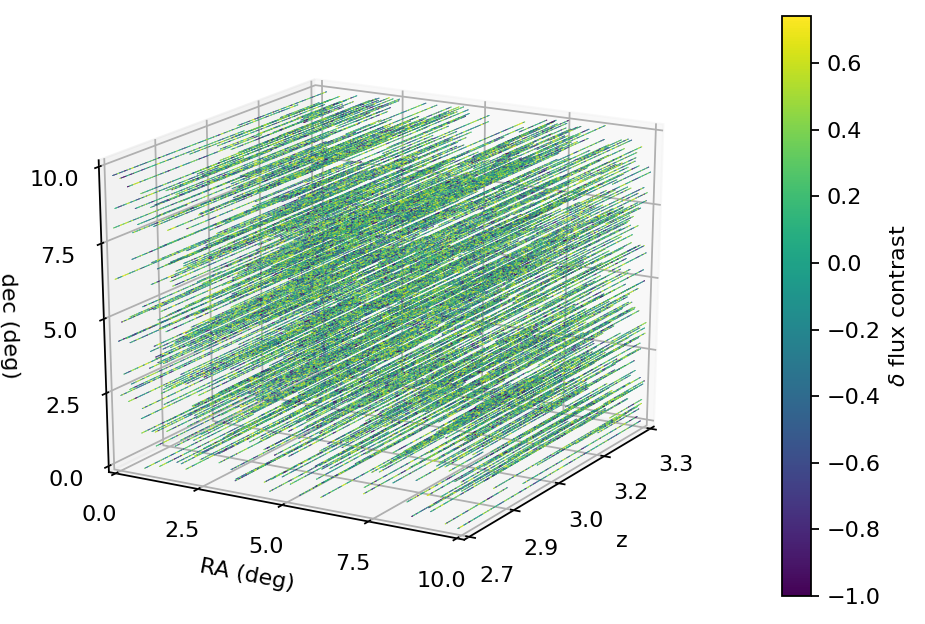}
     \caption{3D simulations of the Ly$\alpha$ forest sampled pixels. \textit{Left:} eBOSS. \textit{Right:} DESI.}
     \label{fig:forest_simulations}
\end{figure*}

To simulate the Ly$\alpha$ forest data of eBOSS/DESI surveys, the hydrodynamic absorption simulation show in Fig. \ref{fig:hydrodynamic_simulations} is replicated four times in a mosaic pattern, as if it was shown in the PAUS simulation.

After this operation, random cells in the simulation array are selected with the quasar density redshift distribution shown in Fig. \ref{fig:quasar_redshift_distribution} (depending on the survey to be simulated), with the redshift of each cell computed as for the Ly$\alpha$ emission simulation. The RA and dec coordinates of the quasar cells are selected randomly from a uniform distribution. The total number of quasar cells (i.e., the number of quasars in the sample) is also computed from the redshift distribution, considering that the simulation has an angular area of 100 deg$^2$ and that only quasars with $z>2.7$ are to be included (since quasars at lower redshift will have all Ly$\alpha$ forests outside the redshift range of the simulation).

The cells between the quasar cells and the observer (the cells in the same angular bins and negative redshift direction) are considered Ly$\alpha$ forest cells, including the quasar cells themselves. Only these forest cells are taken into account for cross-correlation; everything else in the hydrodynamic simulation is masked.

In addition to this, if a quasar is at redshift high enough so that Ly$\beta$ forest appears at $z>2.7$, its forest cells that would be covered by the Ly$\beta$ region are also masked, given that these regions of the quasar spectrum contains both Ly$\alpha$ and Ly$\beta$ absorption lines superimposed from different redshifts. While these Ly$\beta$ forest regions can be used for cross-correlation studies (e.g., \citet{Blomqvist2019a}), here we adopt the conservative approach and remove them from the cross-correlation. These masked Ly$\beta$ cells account for 12\% of the total forest cells.

Regarding the SNR of the forest data, we take as a reference the mean SNR values displayed in \cite{Chabanier2019}, Table 2. These correspond to a high quality sample of the first eBOSS release (selecting 43,751 out of 180,413 visually inspected spectra), and thus are an optimistic estimate of what can be expected in both future eBOSS releases and DESI. For the redshift bin closer to our study ($z\sim3$), the eBOSS data shows $\rm \langle SNR \rangle=6.5$ per forest pixel. However, these values need to be scaled to the bin size of our forest simulation with the following expression, 

\begin{equation}\label{eq:snr_forest_scaling}
{\rm SNR}_{\rm simulation}(\lambda)=\sqrt{\frac{\Delta\lambda_{\rm simulation}}{\lambda \langle R_{\rm eBOSS} \rangle^{-1}}}\langle{\rm SNR}_{\rm eBOSS}\rangle,
\end{equation}

where $\Delta \lambda_{\rm simulation}$ is the wavelength bin size for our forest simulation (determined as in \S\ref{sec:simulation_paus_lyalpha_im}), and $\langle R_{\rm eBOSS} \rangle$ the mean resolution of the eBOSS spectra after reduction with the pipeline used in \cite{Chabanier2019}, which resamples the wavelength pixels of the coadded spectra in logarithmic bins of $\Delta \log_10=10^{-4}$ (or equivalently, R$\sim$4350). This results in a higher SNR, that increases linearly from 9.97 at the lowest redshift to 10.29 at the high redshift end. Gaussian noise is added to each forest cell according to the determined SNR.

A voxel representation of this Ly$\alpha$ forest simulation, displaying only forest cells used for cross-correlation, is shown in Fig. \ref{fig:forest_simulations}, both for eBOSS and DESI expected quasar densities.
 
\section{Simulated cross-correlation estimator}\label{sec:Simulated cross-correlation estimator}

\subsection{Estimator definition}

In order to compute the cross-correlation from the PAUS and eBOSS/DESI simulated datasets explained in the previous section, an estimator of the 2PCF is needed. The estimator used for this work is

\begin{equation}\label{eq:corr_estimator}
\hat{\xi}(r_n)=\frac{\sum_{\rm i} \left( \delta_{\rm i} \sum_{\rm \in Bin(r_{\rm n})} \phi_{\rm j} \right)}{\sum_{\rm i} \left( 1 \sum_{\rm \in Bin(r_{\rm n})} 1\right)}.
\end{equation}

This estimator  is defined for distance bins $r_n$. Since the cells to be cross-correlated have finite volumes, distances are assumed from the coordinates of their centres. Regarding the other terms in the equation, $\delta_{\rm i}$ is the $\delta$ flux of the forest cell $i$, as defined in Eq. (\ref{eq:delta_contrast}), and $\phi_{\rm j}$ is the absolute flux contrast for the pixel $j$ in simulated PAUS images, defined as

\begin{equation}\label{eq:PAUS_flux_contrast}
    \phi_{\rm j} \equiv \frac{F_{\rm \lambda \, j}}{\langle F_{\rm \lambda} \rangle}-1.
\end{equation}

In other words, this estimator is the average value of the products of all cell pairs in a certain distance bin. This distance $r$ in Eq. (\ref{eq:corr_estimator}) is defined as the total distance between cells (monopole cross-correlation), but it could also be defined as the distance projected onto the line of sight (parallel cross-correlation, $\xi(r_\parallel)$), or perpendicular to it (perpendicular cross-correlation, $\xi(r_\bot)$). Consequently, the parallel and perpendicular estimators $\xi(r_{\rm \parallel\, n})$ and $\xi(r_{\rm \bot\, n})$ can be defined simply by switching the definition of distance, $|i-j|$, by $|i-j|\cdot \vec{u}_{\rm los}$ and $|i-j|\times \vec{u}_{\rm los}$ respectively (where $\vec{u}_{\rm los}$ is the unit vector parallel to the line of sight).

Normally, the average computed by this estimator is weighted by a function of the pipeline error, as well as additional errors terms derived from data reduction (e.g. \citet{Font-Ribera2012}). However, for this preliminary work the error in simulated PAUS images is approximately constant, with only slight variations between filters (see Table \ref{tab:sigma_noise_abs}), and the Ly$\alpha$ forest error has been considered negligible, so no weighting has been applied.

The error on the estimator is computed using jackknife resampling. The simulation has been divided in 25 subsamples by imposing uniform cuts in RA and dec. Since space is not sampled uniformly in redshift in this cross-correlation (because Ly$\alpha$ forest available data depends on the quasar redshift distributions), no cuts have been performed in redshift, so all jackknife subsamples cover the whole redshift range of the simulation.

\subsection{Noise bias}

The cross-correlation estimator introduced in  Eq. (\ref{eq:corr_estimator}) is biased if at least one of the signals being cross-correlated contains noise of mean different than zero, which is of particular importance for this study. In order to demonstrate this, let us assume that the estimator is used to cross-correlate two arbitrary observable scalar fields, $f(\textbf{r})$ and  $g(\textbf{r})$. For both fields, a finite number of samples at different points are observed, $f_{\rm i}$ and $g_{\rm i}$,  and from these points the respective means $\langle f \rangle$ and $\langle g \rangle$ are computed. In order to apply the estimator, the contrasts of both fields need to be determined, which, as in Eq. (\ref{eq:delta_contrast}) and Eq. (\ref{eq:PAUS_flux_contrast}), would be done with the following expressions

\begin{equation}\label{eq:fields_contrast}
    f_{\rm contrast\,i}=\frac{f_{\rm i}-\langle f \rangle}{\langle f \rangle}; \quad \quad
    g_{\rm contrast\,i}=\frac{g_{\rm i}-\langle g \rangle}{\langle g \rangle}.
\end{equation}

If $\delta_{\rm i}$ and $\phi_{\rm j}$ are replaced in Eq. (\ref{eq:corr_estimator}) by $f_{\rm contrast\,i}$ and $g_{\rm contrast\,j}$, and these are substituted by its definition in  Eq. (\ref{eq:fields_contrast}), the following expression can be obtained

\begin{equation}\label{eq:corr_estimator_developed}
\xi(r)=\frac{\sum_{\rm i} \left[ (f_{\rm i}-\langle f \rangle) \sum_{\rm j}^{\rm r} (g_{\rm j}-\langle g \rangle) \right]}{\langle f \rangle \langle g \rangle \sum_{\rm i} \left( 1 \sum_{\rm j}^{\rm r} 1\right)}.
\end{equation}

Here the second summation in the right side of Eq. (\ref{eq:corr_estimator}) has been rewritten as $\sum_{\rm j}^{\rm r}$ for simplicity. Now, let us consider that the field $g(\textbf{r})$ is the sum of two independent fields, the signal $S(\textbf{r})$ and the noise $N(\textbf{r})$, so

\begin{equation}\label{eq:field_signal+noise}
g(\textbf{r})=S(\textbf{r})+N(\textbf{r}).
\end{equation}

By our definition, the noise $N(\textbf{r})$ is uncorrelated with $f(\textbf{r})$, so for a sample large enough a hypothetical estimated cross-correlation between $f(\textbf{r})$ and $N(\textbf{r})$ would tend to zero. Following Eq. (\ref{eq:corr_estimator_developed}), this can be expressed as

\begin{equation}\label{eq:null_correlation_noise}
\sum_{\rm i} \left[ (f_{\rm i}-\langle f \rangle) \sum_{\rm j}^{\rm r} (N_{\rm j}-\langle N \rangle) \right] \to 0.
\end{equation}

Conversely, the hypothetical cross-correlation $\xi_S(r)$ between $f(\textbf{r})$ and $S(\textbf{r})$ would be

\begin{equation}\label{eq:correlation_signal}
\xi_S(r)=\frac{\sum_{\rm i}\left[(f_{\rm i}-\langle f \rangle) \sum_{\rm j}^{\rm r} (S_{\rm j}-\langle S \rangle) \right]}{\langle f \rangle \langle S \rangle \sum_{\rm i} \left( 1 \sum_{\rm j}^{\rm r} 1\right)}.
\end{equation}

Nevertheless, only the field $g(\textbf{r})$ can be observed, and thus the only cross-correlation that can be computed is that of the $f(\textbf{r})$ with $S(\textbf{r})$ plus $N(\textbf{r})$:

\begin{equation}\label{eq:correlation_s+n}
\xi_{\rm S+N}(r)=\frac{\sum_{\rm i} \left\{ (f_{\rm i}-\langle f \rangle) \left[ \sum_{\rm j}^{} (S_j-\langle S  \rangle)+\sum_{\rm j}^{} (N_j-\langle N \rangle) \right]\right\}}{\langle f \rangle \langle S + N \rangle \sum_{\rm i} \left( 1 \sum_{\rm j}^{} 1\right)}.
\end{equation}

If a sample large enough is assumed, Eq. (\ref{eq:null_correlation_noise}) holds true, and since the noise component of the cross-correlation tends to zero, the denominator in Eq. (\ref{eq:correlation_signal}) and Eq. (\ref{eq:correlation_s+n}) is identical. Therefore, the following relation can be derived between the hypothetical cross-correlation of the signal, $\xi_{\rm S}(r)$, and the actual cross-correlation of the signal with noise, $\xi_{\rm S+N}(r)$, is

\begin{equation}\label{eq:noise_bias}
\xi_{\rm S+N}(r)=\frac{\langle S \rangle}{\langle S+N \rangle}\xi_S(r).
\end{equation}

If the noise of the observable $g(\textbf{r})$ had mean zero, we would have $\xi_{\rm S+N}(r)=\xi_{\rm S}(r$), and thus the estimator would be unbiased. However if we consider PAUS images to be the observable $g(\textbf{r})$, the noise $N(\textbf{r})$ would be the foregrounds plus instrumental noise. The first component necessarily has a mean larger than zero, since it is a sum of observed fluxes, while the second also should in principle, given that it includes effects such as scattered light and airglow, which are strictly positive.

Nevertheless, this noise bias does not affect the SNR, and thus the probability of detection. Considering that the error is computed via jackknife resampling (i.e., the $\sigma$ of the cross-correlation computed for different subsamples), this noise bias will multiply the cross-correlation value and its error equally, and therefore will cancel out when computing the SNR.

\section{Theoretical correlation function}\label{sec:Theoretical correlation function}

To validate the result of the simulated cross-correlation, as well as to derive the clustering, comparison against a theoretical 2PCF is needed. The first step is to compute the unbiased matter-matter 2PCF from the theoretical matter power spectrum. For this work, this 2PCF has been initially computed as a field depending on two variables, the distances parallel and perpendicular to the line of sight, $r_{\rm \parallel}$ and $r_{\rm \bot}$ respectively, with the following expression (e.g. see \citealt{Hui2007, Gazta2009})

\begin{equation}\label{eq:theoretical_correlation}
\xi(r_{\rm \parallel}, r_{\rm \bot})=\frac{1}{2 \pi^2}  \int_0^{\infty} dk \, k P_{\rm nl}(k) \frac{\sin \left( k\sqrt{r_{\rm \parallel}^2+r_{\rm \bot}^2} \right)}{\sqrt{r_{\rm \parallel}^2+r_{\rm \bot}^2}}\exp(-kr_{\rm cut}).
\end{equation}

Where $P_{\rm nl}(k)$ is the non-linear matter power spectrum computed with \textsc{camb} \citep{Lewis2000}, and the non-linear modelling of \textsc{halofit} \citep{Peacock2014}. This power spectrum has been computed at the redshift of the hydrodynamic simulation snapshot ($z=3$), using its cosmology. Regarding other terms, $r_{\rm cut}$ is the radius of the exponential cutoff set in order to avoid large oscillations in the theoretical 2PCF due to small-scale effects that are not represented in its counterpart measured in the simulation. For this study, the chosen value for this cutoff is $r_{\rm cut}=3$ Mpc/h.

By definition, there is no anisotropy in Eq. (\ref{eq:theoretical_correlation}), which may make the computation of the 2PCF in two directions seem redundant. Nevertheless, two effects that are to be taken into account will break the isotropy of the function: the smoothing introduced by the binning of the simulated data, and the effect of redshift-space distortions (RSDs).

\subsection{Smoothing}\label{sec:smoothing}

This effect arises from the fact that correlation is being performed between spatial cells with finite volumes, whose value of the field to cross-correlate is the average over the volume of the cell. If the length of these cells to cross-correlate is equal or smaller than the binning of the correlation estimator (Eq. (\ref{eq:corr_estimator})), this effect will be negligible, given that by binning the estimator already averages over a similar length. This is the case for the Ly$\alpha$ forest cells in all directions or the PAUS cells in RA and dec directions, where their length (1.56 Mpc/h, given by the hydrodynamic simulation bins) is smaller than the binning of the estimated cross-correlation (see Fig. \ref{fig:simulated_cross-correlation_fg}).

On the other hand, this effect is not negligible for the redshift direction in PAUS cells, where the mean cell size is 56.25 Mpc/h (since redshift bins have been merged to simulate PAUS filters). Averaging the Ly$\alpha$ flux in PAUS images over such distances will certainly have an effect on the estimated cross-correlation, which also has to be simulated in the theoretical 2PCF. Considering that the redshift direction in the simulation has a direct correspondence with $r_{\rm \parallel}$ in Eq. (\ref{eq:theoretical_correlation}), this smoothing can be emulated by averaging each point in the computed $\xi(r_{\rm \parallel}, r_{\rm \bot})$ field over a length in $r_{\rm \parallel}$ equal to the average PAUS cell size

\begin{equation}\label{eq:smoothing_correlation}
\bar{\xi}(r_{\rm \parallel}, r_{\rm \bot})=\frac{1}{l_{\rm \parallel}} \int_{\rm r_{\rm \parallel}-l_{\rm \parallel}/2}^{r_{\rm \parallel}+l_{\rm \parallel}/2}dr'_{\rm \parallel}\, \xi(r'_{\rm \parallel}, r_{\rm \bot}).
\end{equation}

Where $l_{\rm \parallel}=56.25$ Mpc/h. By definition of the 2PCF, $r>0$, so for $r_{\rm \parallel}<l_{\rm \parallel}/2$ this expression changes to

\begin{equation}\label{eq:smoothing_correlation2}
\bar{\xi}(r_{\rm \parallel}, r_{\rm \bot})=\frac{1}{l_{\rm \parallel}} \left[ \int_{\rm 0}^{r_{\rm \parallel}+l_{\rm \parallel}/2}dr'_{\rm \parallel}\, \xi(r'_{\rm \parallel}, r_{\rm \bot})+\int_{\rm 0}^{l_{\rm \parallel}/2-r_{\rm \parallel}}dr'_{\rm \parallel}\, \xi(r'_{\rm \parallel}, r_{\rm \bot}) \right].
\end{equation} 

\begin{figure}\centering
 	\includegraphics[width=\columnwidth]{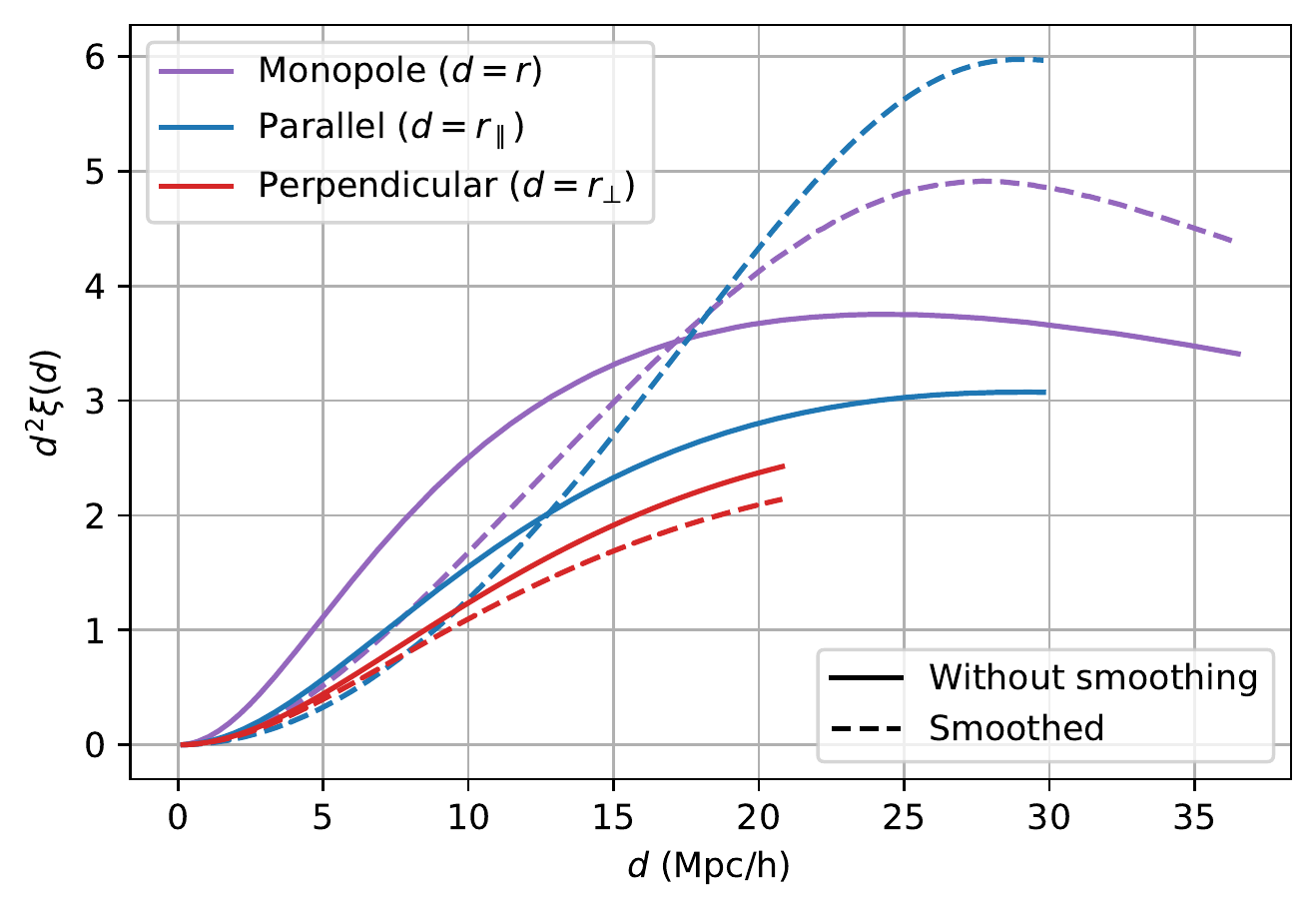}
     \caption{Theoretical unbiased 2PCFs times squared distance, before and after applying smoothing. The distance range of each one of the 2PCFs is the same as all the results shown in \S \ref{sec:Results}. Redshift space distortions have not been considered in this figure.}
     \label{fig:smoothing_comparison}
\end{figure}

If the 2PCF is interpreted as an average product of cell pairs at a certain distance, such as in the estimator, this last expression represents the case where the small Ly$\alpha$ forest cell lies inside the redshift range of the PAUS cell it is being cross-correlated with. The smoothing integral needs to cover the whole $l_{\rm \parallel}$, but since the distance between cells necessarily has to be non-negative, the integral is truncated in two terms: one for the portion of the PAUS cell at higher redshift than the Ly$\alpha$ forest cell, and another for the portion at lower redshift. Fig. \ref{fig:smoothing_comparison} shows the effect of this 2PCF smoothing (dashed lines) compared to the non-smoothed 2PCF (solid lines) for the three correlation types considered in this work.

\subsection{Monopole, parallel and perpendicular 2PCF}\label{sec:monopole_parallel_and_perpendicular_2pcf}

This two-dimensional 2PCF has been converted to a 2PCF depending solely on a single distance parameter, either the total distance between cell pairs $r=\sqrt{r_{\rm \parallel}^2+r_{\rm \bot}^2}$, or the parallel/perpendicular distances, in order to be compared to the estimator defined in Eq. (\ref{eq:corr_estimator}). The estimator could also be defined as a function of both $r_\parallel$ and $r_\bot$, however, this would greatly reduce the number of cell pairs available per bin, and thus the SNR of the measured cross-correlation.

For the monopole 2PCF, this has been performed by computing $\bar{\xi}_{\rm mm}(r_{\rm \parallel}, r_{\rm \bot})$ in a very fine uniform grid of $r_{\rm \parallel}$, $r_{\rm \bot}$ values, and then averaging these values in bins of total distance $r=\sqrt{r_{\rm \parallel}^2+r_{\rm \bot}^2}$. Regarding the parallel and perpendicular 2PCFs, they have been obtained from the theoretical two-dimensional 2PCF simply by numerical integration, according to the following expressions

\begin{equation}\label{eq:parallel-perpendicular_corr_theoretical}
\begin{split}
\bar{\xi}_{\rm mm}(r_\parallel)=\frac{1}{R_\bot}\int_0^{R_\bot} dr_\bot \, \xi_{\rm mm}(r_\parallel,r_\bot) \\
\bar{\xi}_{\rm mm}(r_\bot)=\frac{1}{R_\parallel}\int_0^{R_\parallel} dr_\parallel \, \xi_{\rm mm}(r_\parallel,r_\bot) ,
\end{split}
\end{equation}

where $R_\parallel$ and $R_\bot$ are the maximum binning distances used by the estimator in Eq. (\ref{eq:corr_estimator}) for the parallel and perpendicular directions. These 2PCFs, unlike the monopole, depend on the total range over which the correlation is computed, which makes them less suitable for comparison of the results against the theory. Consequently, only the monopole 2PCF will be used to compare the results of the simulation against the theory in \S \ref{sec:Results}.

\subsection{Bias and RSDs}\label{sec:bias_and_rsds}

In addition to this smoothing effect, bias from the tracers also needs to be taken into account, as well as the effect of RSDs. Since the scales studied in this work are large enough, the only RSD effect considered is the Kaiser effect \citep{Kaiser1987}.

So far, the unbiased matter-matter 2PCF has been considered (called $\bar{\xi}_{\rm mm}$ henceforth), but the cross-correlation in this work uses Ly$\alpha$ emission and Ly$\alpha$ forest absorption. The power spectrum of a tracer $t$ correlated with itself can be obtained from the unbiased 2PCF with the following expression


\begin{equation}\label{eq:bias_correlation}
P_{\rm tt}(k)=b_{\rm tt}^2(1+\beta_{\rm tt}{\mu_{\vec{k}}}^2)^2 P(k).
\end{equation}

Where $b_{\rm tt}$ is the bias of the tracer (Ly$\alpha$ emission and Ly$\alpha$ forest in this case), $\mu_k$ is the cosine of the angle between the vector position vector in Fourier space $\vec{k}$ and the line of sight, and $\beta_{\rm tt}$ is the effective RSD parameter. $P_{\rm tt}(k)$ is the power spectrum of the tracer; given that RSDs are included in the expression, it is implied to be in redshift space.

If we are to compute the 2PCF from this $P_{\rm tt}(k)$, we need to take into account the dependency of $\mu_{\vec{k}}$ with $r_\parallel$ and $r_\bot$.  For the two-dimensional 2PCF $\mu_{\vec{k}}$ will depend on the values of $r_\parallel$ and $r_\bot$, but for the monopole, parallel and perpendicular correlation, this dependency vanishes. For the parallel and perpendicular case, $\mu_{\vec{k}}=1$ and $\mu_{\vec{k}}=0$ respectively, while for the monopole, after integration in all directions Eq. (\ref{eq:bias_correlation}) becomes

\begin{equation}\label{eq:rsd_monopole}
\bar{\xi}_{\rm tt}(r)=b_{\rm tt}^2\left(1+\frac{2}{3}\beta_{\rm tt}+\frac{1}{5}\beta_{\rm tt}^2\right)^2\bar{\xi}_{\rm mm}(r).
\end{equation}
For this work, two tracers are considered: the Ly$\alpha$ emission (denoted by $\rm e$) and the Ly$\alpha$ absorption that generates the Ly$\alpha$ forest (denoted by $\rm a$); the autocorrelation of both tracers will be computed to validate the simulation, as explained in \S\ref{sec:comparison_against_theory}. The assumed bias factors $b$ for both tracers are already explained in \S\ref{sec:independent_simulations}, while the $\beta$ parameter needs different assumptions for each case.

Given that the Ly$\alpha$ emission has been considered proportional to the square of the matter density field, its $\beta$ will be the same as the matter density, scaled by the bias factor $b_{aa}$. Following the approximation in \cite{Kaiser1987} for linear theory, we find

\begin{equation}\label{eq:rsd_emission}
\beta_{\rm ee}=\frac{\Omega_m(z)^{0.6}}{b_{\rm ee}}.
\end{equation}Regarding the Ly$\alpha$ absorption, $\beta_{\rm aa}$ is independent from $b_{\rm aa}$, since the nonlinear transformation applied to obtain the Ly$\alpha$ absorption field (Eq. (\ref{eq:delta_contrast})) does not preserve the flux between real and redshift space. According to \cite{Slosar2011}, the value of $\beta_{\rm aa}$ for different simulations depends on their resolution, with values oscillating between 1 and 1.5 at z$\sim$2.25. Recent observations of BAO with the Ly$\alpha$ forest autocorrelation yield $\beta_{\rm aa}\sim1.8$ at $z=2.34$ \citep{Agathe2019}, but also show some evidence of $\beta_{\rm aa}$ decreasing with redshift. Since the mean redshift of this work ($z\sim3$) is higher than any of the cited values of $\beta_{\rm aa}$, and the hydrodynamic simulation is relatively low-resolution, we decide to adopt the conservative value of $\beta_{\rm aa}=1$.

So far, we have discussed how to model the effects of the bias and the RSDs for the autocorrelation of the Ly$\alpha$ emission and the Ly$\alpha$ absorption from the forest; however, the cross-correlation between both tracers also needs to be modeled. In order to do so, we compute the effective biases of each tracer for cross-correlation, $\hat{b_{\rm t}}$ as the tracer bias times the square root of its RSD factor. For example, for the monopole, following Eq. (\ref{eq:rsd_monopole}) the effective bias would be

\begin{equation}\label{eq:effective_bias}
\hat{b}_{\rm t}=b_{\rm tt}\left(1+\frac{2}{3}\beta_{\rm tt}+\frac{1}{5}\beta_{\rm tt}^2\right).
\end{equation}Therefore, the monopole cross-correlation can be expressed as

\begin{equation}\label{eq:cross-correlation_prediction}
\hat{\xi}_{\rm ea}(r)\simeq -\hat{b}_a(r)\hat{b}_e(r)\bar{\xi}_{\rm mm}(r).
\end{equation}
Where the minus sign comes from the fact that the cross-correlation is between an emission and an absorption field. Table \ref{tab:bias_table} sums up the $b_{\rm tt}$ and $\beta_{\rm tt}$ considered for both Ly$\alpha$ emission and absorption, as well as the resulting effective bias $\hat{b_{\rm t}}$ for the monopole.

\begin{table}
\centering
	\caption{Bias factor $b_{\rm tt}$ and RSD parameter $\beta_{\rm tt}$ considered for Ly$\alpha$ emission and absorption, as well as the resulting effective bias for cross-correlation $\hat{b_{\rm t}}$, according to Eq. \ref{eq:effective_bias}.}
 	\label{tab:bias_table}
 	\begin{tabular}{rccc}
 		\hline
 		 & $b_{\rm tt}$ & $\beta_{\rm tt}$ & $\hat{b}_{\rm t}$ monopole\\
 		\hline
 		Ly$\alpha$ emission  & 2.000 & 0.488 & 2.343\\
        Ly$\alpha$ absorption  & 0.336 & 1.000 & 0.557\\
		\hline
 	\end{tabular}
 \end{table}

\section{Results}\label{sec:Results}

The results presented in this paper are divided between three subsections. In Section \ref{sec:comparison_against_theory}, the effective bias of the tracers of the simulation (with RSDs included), as defined in Eq. (\ref{eq:effective_bias}) is measured by comparing the absorption and emission auto-correlations against the theoretical 2PCFs. These measured effective biases are then applied to the theoretical 2PCF, and compared against the estimated cross-correlation  in simulations without either foregrounds or PAUS noise (only Ly$\alpha$ emission in the simulated images). In Section \ref{sec:detection_probability_uncorrelated}, the probability of a detection (SNR>3) when adding foregrounds and instrumental noise (using the uncorrelated extrapolation) is explored at different scales. Four cases of cross-correlation are explored: PAUS-eBOSS, PAUS-DESI, and the cross-correlation of DESI with two hypothetical extensions of PAUS (an increase in exposure time, PAUS deep, and in survey field, PAUS extended). Finally, in Section \ref{sec:detection_probability}, we summarise the same results, but using the correlated noise directly measured from PAUS images, instead of the uncorrelated extrapolation.

\subsection{Cross-correlation without noise or foregrounds. Comparison against theory}\label{sec:comparison_against_theory}

In order to compare the cross-correlation results against the theoretical prediction (and thus validate that the cross-correlation results are sound), the actual effective biases of the tracers of the hydrodynamic simulation need to be measured and compared against the expected values from Table \ref{tab:bias_table}.

These effective biases of the simulation have been measured by correlating the emission/absorption arrays of the hydrodynamic simulation (Fig. \ref{fig:hydrodynamic_simulations}) with themselves, using the same binning as in the PAUS-eBOSS/DESI simulation (wide redshift bins for PAUS, only Ly$\alpha$ forest cells for eBOSS/DESI). No foregrounds or noise were added for this correlation, since they do not have the same physical units, and the purpose of this calculation is just to determine the real bias while testing that the binning of the simulation and the smoothing effect are properly taken into account. Considering Eq. (\ref{eq:rsd_monopole}) and Eq. (\ref{eq:effective_bias}), the effective bias of the tracer can be estimated from the smoothed theoretical prediction $\bar{\xi}_{\rm mm}$ and the estimated correlation of the tracer $\hat{\xi}_{\rm tt}$ with

\begin{equation}\label{eq:bias_recovery}
\hat{b}_t(r)=\sqrt{\frac{\hat{\xi}_{\rm tt}(r)}{\bar{\xi}_{\rm mm}(r)}}.
\end{equation}

\begin{figure*}
 	\includegraphics[width=\columnwidth]{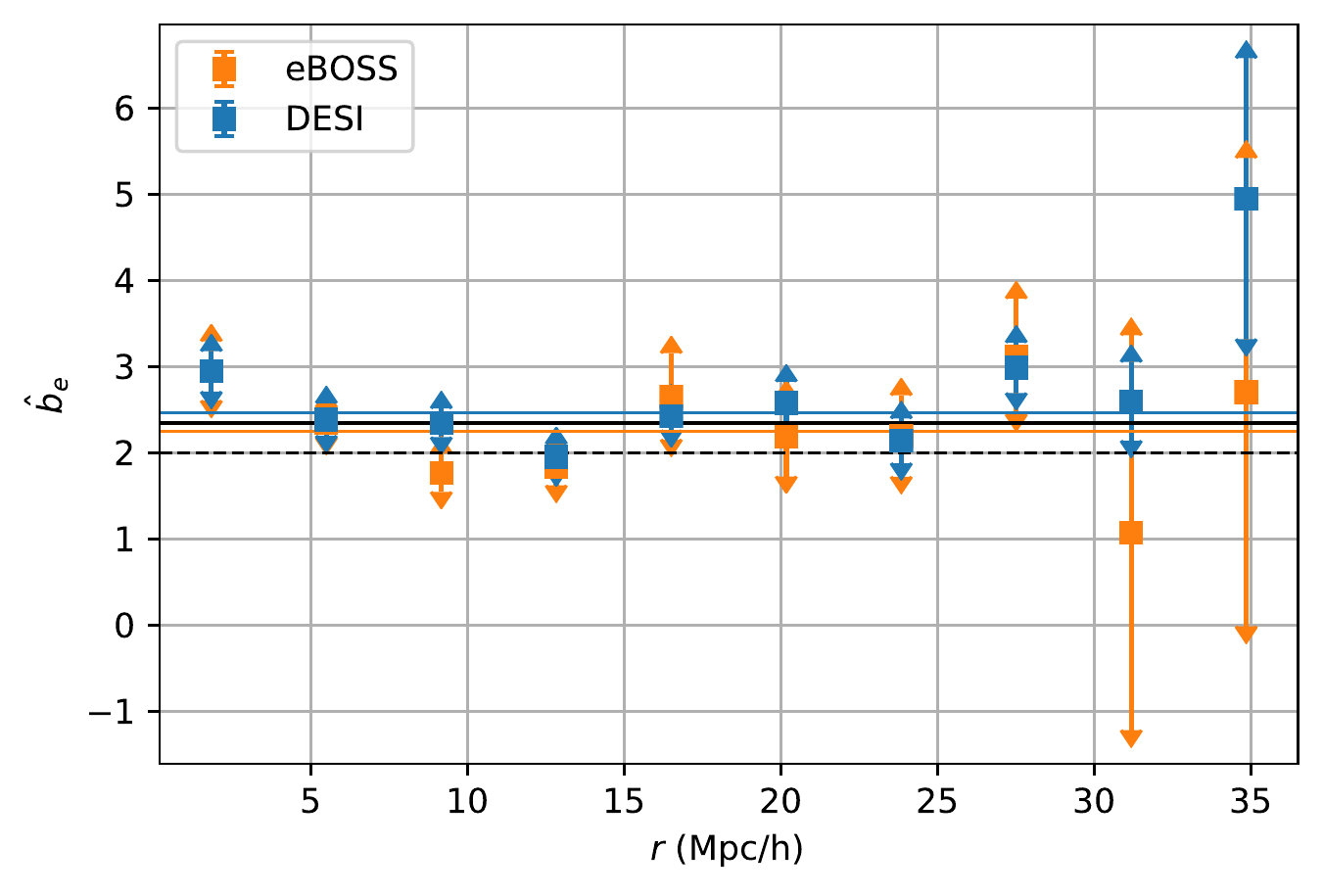}
 	\includegraphics[width=\columnwidth]{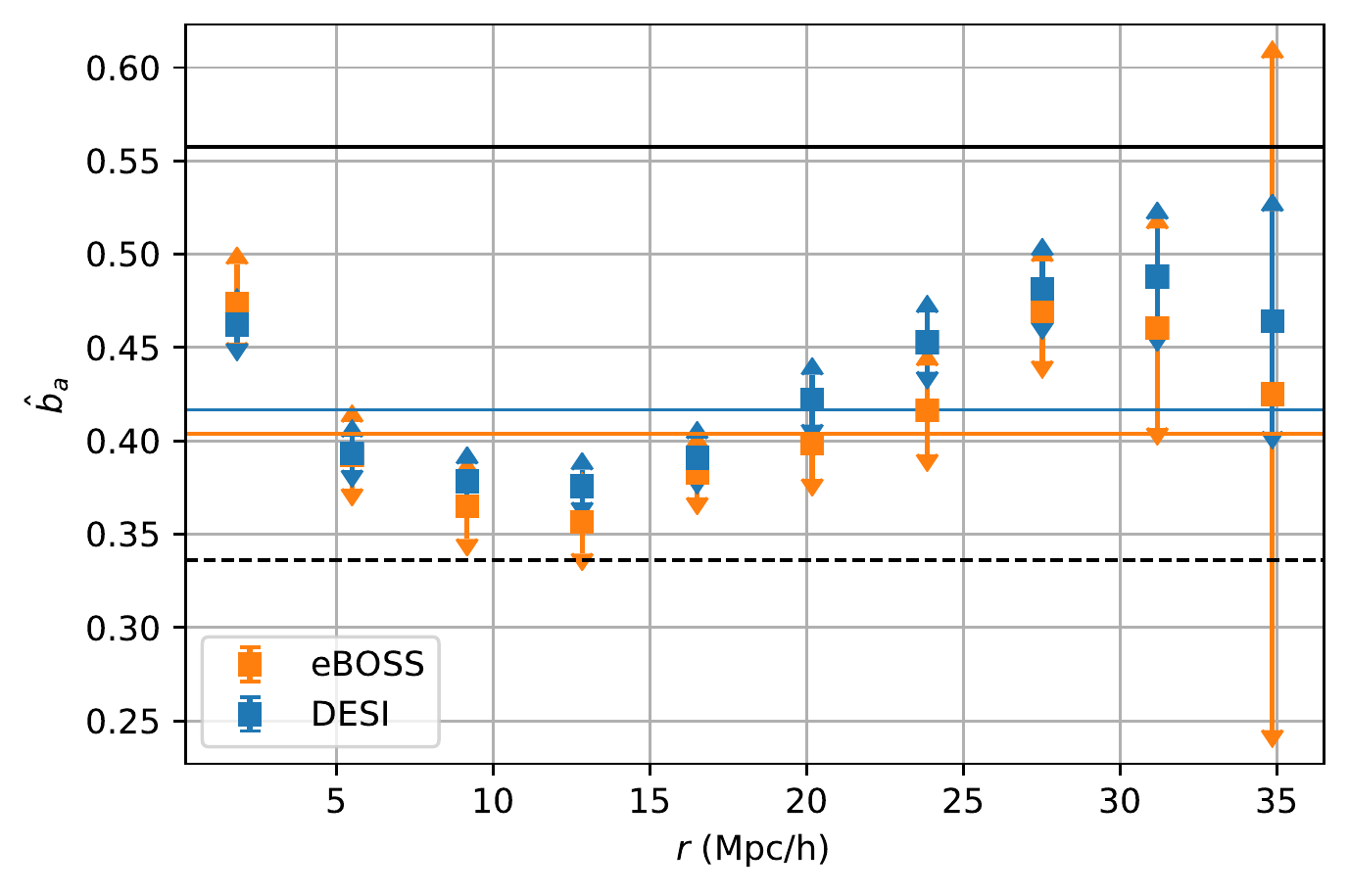}
     \caption{Emission bias (left) and absorption bias (right), measured as described in Eq. (\ref{eq:bias_recovery}), for correlations using eBOSS and DESI Ly$\alpha$ forest binnings. Solid black line represents the approximate effective bias from \S \ref{sec:hydrodynamic_simulation}; dashed black line represent the theoretical bias without considering RSDs ($\beta=0$). Coloured horizontal lines are the weighted average of the recovered effective bias for the respective surveys.}
     \label{fig:measured_biases}
\end{figure*}

Where $t$ is any tracer, and the expression has been considered only for the monopole 2PCF.  The results of this bias determination can be seen in Fig. \ref{fig:measured_biases}. The error of the bias at all distance bins is simply the propagated error of the cross-correlation; any error that could be included in the theoretical 2PCF (e.g., cosmic variance) has been considered negligible.

As can be seen in Fig. \ref{fig:measured_biases}, the actual measured $\hat{b}_{\rm a}$, around 0.4, is smaller by $\sim$25\% than the expected value from Table \ref{tab:bias_table}, but for the emission field $\hat{b}_{\rm e}$ is actually really close to the predicted value. This is to be expected, the Ly$\alpha$ emission field is proportional to the square of the matter density field, which already gives an exact value of the bias, and the effect of RSDs in this case are well described by linear theory. However, for the Ly$\alpha$ forest both $b$ and $\beta$ are uncertain (especially the latter), and the reference values available come from measurements/simulations at lower redshifts, so such discrepancies are reasonable.

With these measured biases, the simulated cross-correlation can be compared to the theoretical 2PCF by applying Eq. (\ref{eq:cross-correlation_prediction}), but instead of using the effective bias values of Table \ref{tab:bias_table}, we apply the measured effective biases from Fig. \ref{fig:measured_biases} to each distance bin.

\begin{figure*}
 	\includegraphics[width=\columnwidth]{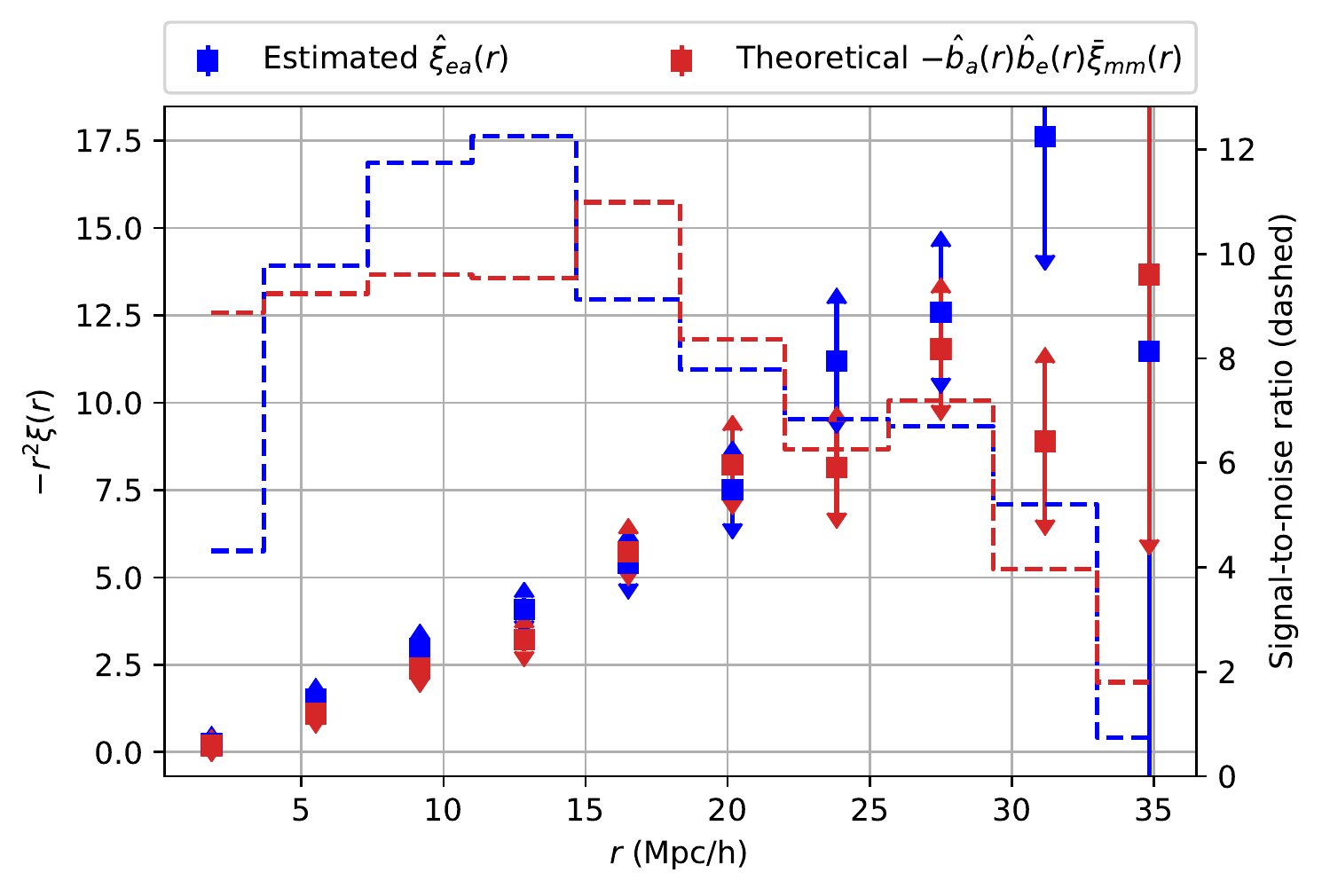}
 	\includegraphics[width=\columnwidth]{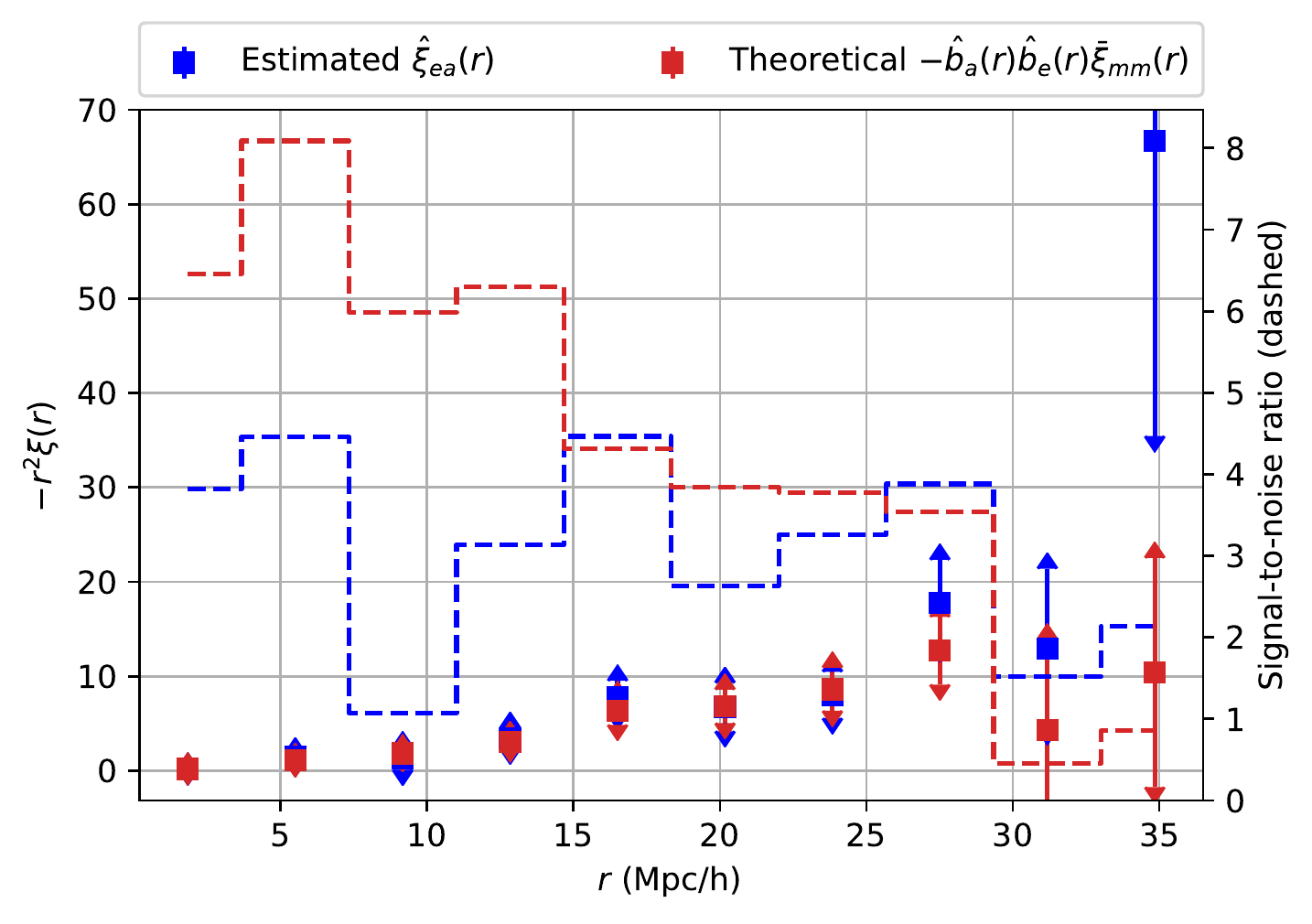}
     \caption{Comparison of the simulated cross-correlation with foregrounds, without instrumental noise, and the theoretical 2PCF with the measured biases. The points with errorbars represent the cross-correlation values (left y axis), while the dashed line represents the SNR of each distance bin (right y axis). \textit{Left panel:} PAUS-DESI. \textit{Right panel:} PAUS-eBOSS}.
     \label{fig:simulated_cross-correlation_fg}
\end{figure*}

A comparison of the simulated cross-correlation, without either foregrounds or instrumental noise, to the theoretical 2PCF with the measured biases is displayed in Fig \ref{fig:simulated_cross-correlation_fg}. Only the monopole 2PCF is displayed, since the parallel and perpendicular 2PCF depend on the range in which the 2PCF is computed, as shown in Eq. \ref{eq:parallel-perpendicular_corr_theoretical}. No foregrounds or instrumental noise have been added both to ensure a good SNR to validate our model, and because the noise bias described in Eq. \ref{eq:noise_bias} would also need to be corrected to compare the simulation against theory.

With no foregrounds or instrumental noise, there is a clear detection of cross-correlation at $r>30$ Mpc/h with DESI, and several bins show a clear detection up to $r\sim 30$ Mpc/h with eBOSS. For all the bins with a detection, the errorbars of the theoretical prediction and the actual cross-correlation overlap; this validates the simulated cross-correlation. Besides, this also proves that, for an ideal case without any other sources of noise, this cross-correlation could be used to constrain either the bias of the tracers or the 2PCF on scales up to $\sim$30 Mpc/h.

Nevertheless, when the foregrounds and the instrumental noise from PAUS are added to the simulation, the general SNR of the cross-correlation drops greatly. Therefore, instead of simulating the cross-correlation and comparing to the theory (assuming that a detection is almost certain), a different approach has been taken to evaluate the probability of a detection.

\subsection{Cross-correlation with uncorrelated noise: Probability of detection}\label{sec:detection_probability_uncorrelated}

\subsubsection{PAUS-eBOSS/DESI}\label{sec:real_detection_probability_uncorrelated}

As explained in \S \ref{sec:Simulation of the survey data}, a simulation of the cross-correlation contains three stochastic elements: the instrumental noise in PAUS images, the Gaussian noise inserted in the Ly$\alpha$ forest simulation, and the quasar cells in eBOSS/DESI  that determine the Ly$\alpha$ forest cells to be sampled (following the redshift distributions in Fig. \ref{fig:quasar_redshift_distribution}). Nevertheless, the noise in the Ly$\alpha$ forest is clearly subdominant (its addition does not alter the results), so we will discuss only the PAUS noise and the quasar positions from now on.

Without the instrumental noise, different realisations of the Ly$\alpha$ forest quasar positions do not modify significantly the cross-correlation results. Nevertheless, when the instrumental noise (using the uncorrelated extrapolation) is added to the PAUS simulation, the SNR of the cross-correlation heavily decreases, up to the point of a detection (SNR>3) depending on the realisation of the noise and the Ly$\alpha$ forest (i.e., the SNR is not consistent between different runs of the simulation pipeline). Fixing one of these stochastic elements (either the Ly$\alpha$ forest position or the instrumental noise) does not gives consistent results either.

Therefore, the approach we have taken is to simulate the cross-correlation 1000 times, with different realisations of the instrumental noise and the Ly$\alpha$ forest quasars each time, and compute the probability of detection (SNR>3). For each one of the realisations, the monopole, parallel and perpendicular 2PCF have been computed using 12 uniform distance bins; finer distance bins would result in empty bins (without any cell pairs) for some cases. 

In addition to this, the cross-correlation has also been computed another 1000 times for each case, but with the Ly$\alpha$ emission in PAUS images mirrored both in the RA and dec axes. This way, the actual cross-correlation between the simulated PAUS images and the Ly$\alpha$ forest should be null, as one of the two signals has been inverted, so any detection that results from this cross-correlation is inherently spurious. In fact, the 2PCF has only been computed with 12 uniform bins partially because applying other uniform binnings seemed to increase the spurious detections without a larger increase in real ones.

\begin{figure}
 	\includegraphics[width=\columnwidth]{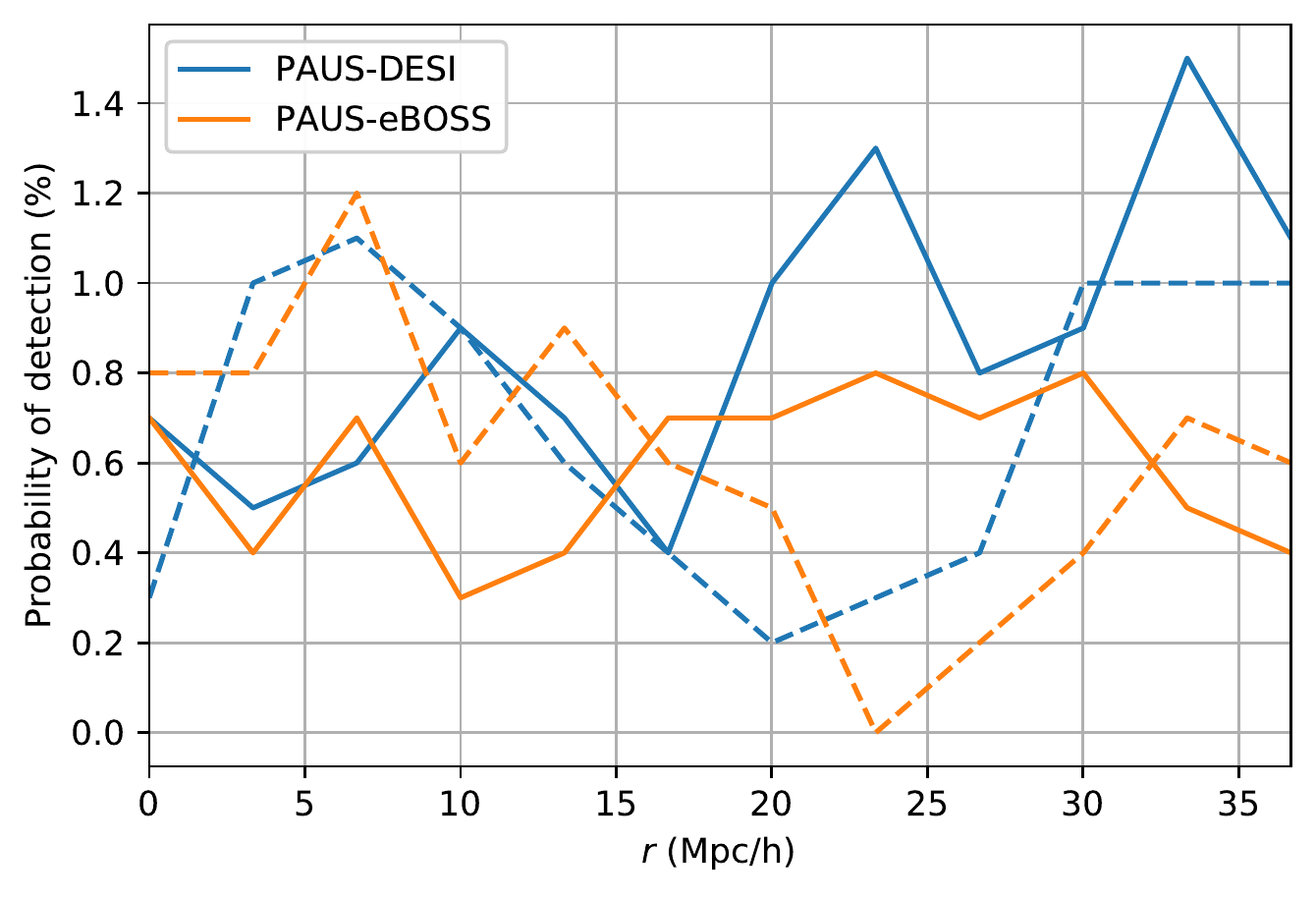}
 	\includegraphics[width=\columnwidth]{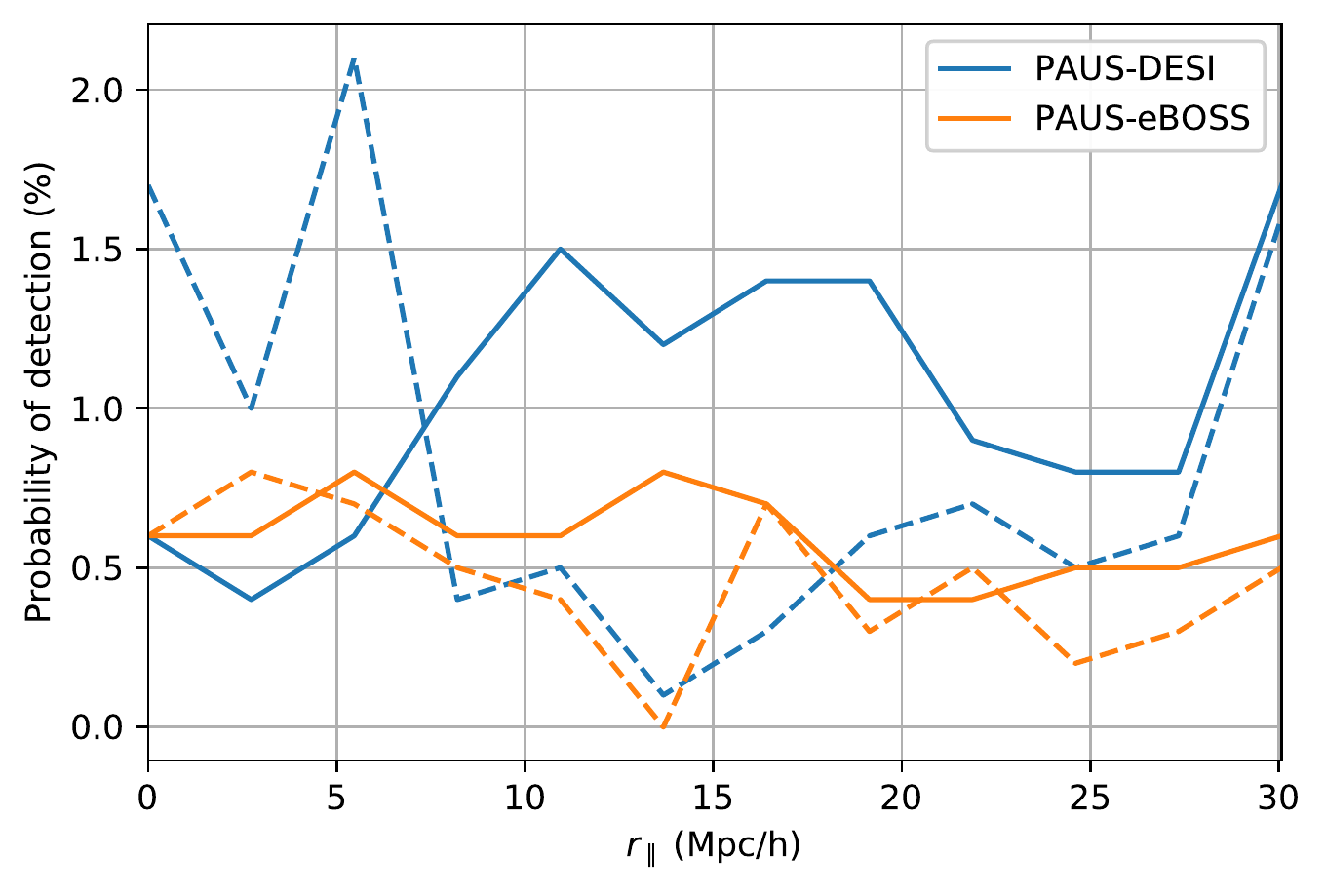}
     	\includegraphics[width=\columnwidth]{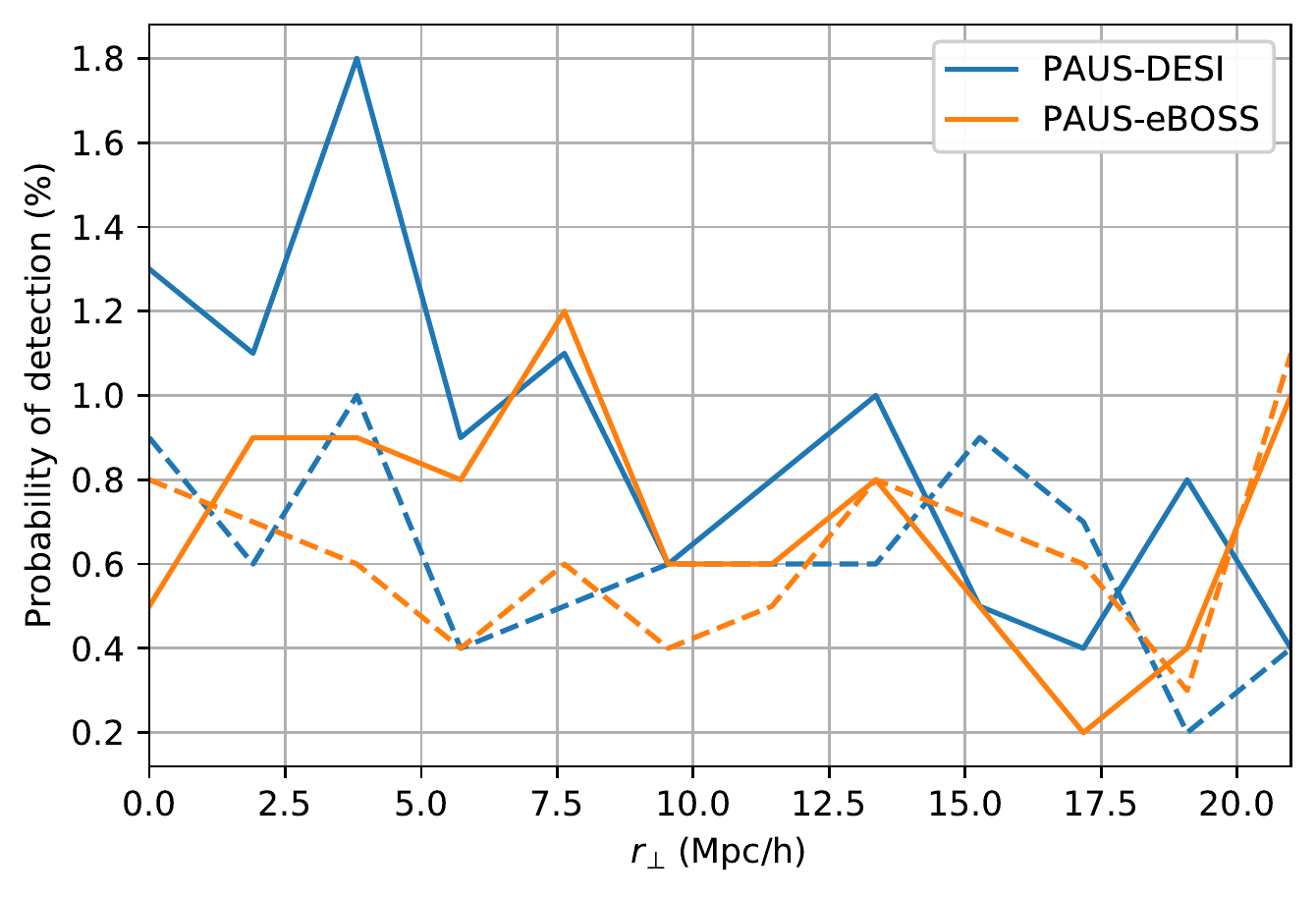}
     \caption{Probability of a detection as a function of distance in the simulated cross-correlation PAUS-eBOSS and PAUS-DESI, for 1000 different realisations of instrumental noise+Ly$\alpha$ forest. Solid line displays the actual probability of any detection, dashed line shows the probability of a spurious detection. \textit{Top panel:} Monopole 2PCF. \textit{Middle panel:} Parallel 2PCF. \textit{Bottom panel:} Perpendicular 2PCF.}
     \label{fig:eBOSS_DESI_probability}
\end{figure}

Fig. \ref{fig:eBOSS_DESI_probability} displays the probability of detection for these 1000 runs for PAUS-eBOSS and PAUS-DESI: the solid line represents the probability of any detection from the real cross-correlation (this includes the spurious ones), and the dashed line the probability of a spurious detection (for the cross-correlation with inverted Ly$\alpha$ emission).

As it would be expected, cross-correlation with DESI noticeably increases the detection probability for all three 2PCFs; however, the probability of any detection is still very small. In the PAUS-eBOSS case, the solid line is barely above the dashed one, which implies the the probability of a detection is almost negligible. In PAUS-DESI, there are some regions where the probability of any detection is clearly above the dashed line; these are the cases that we will discuss.

First, the monopole/parallel 2PCF and the perpendicular 2PCF seem to sample better different scales: the parallel 2PCF has a lower probability of detection at small scales, and shows an increase in detection probability around 10 Mpc/h, while the perpendicular 2PCF has exactly the opposite behaviour. The cause of these contrasting trends in the detection probability for different 2PCFs is the smoothing effect that PAUS filters have in the parallel (redshift) direction, displayed in Fig. \ref{fig:smoothing_comparison}.

In the parallel cross-correlation, and to a lesser extent, the monopole 2PCF, smoothing decreases the absolute value of the 2PCF at scales of 10 Mpc/h and 15 Mpc/h respectively, while at larger scales the 2PCFs are increased (at least, as far as the size of the hydrodynamic simulation allows to compute the 2PCF, 30-35 Mpc/h). This trend matches almost perfectly the detection probabilities in Fig. \ref{fig:eBOSS_DESI_probability}, with sharp increases in the monopole and parallel 2PCF at the same scales.

On the other hand, the perpendicular 2PCF in Fig. \ref{fig:smoothing_comparison} shows a smaller decrease, even when going to larger scales than the ones depicted in Fig. \ref{fig:smoothing_comparison}; this small effect of the smoothing results in higher detection probabilities at smaller scales, where the 2PCF has higher absolute values. This result shows that the parallel and perpendicular 2PCFs are highly complementary, and both should be taken into account for any future observational studies of Ly$\alpha$ IM with PAUS (or similar surveys) in order to maximise the probability of a detection at all scales.

Nevertheless, it is worth noting that, even in these regions where the detection probability increases, the difference with the spurious detection probability is really modest, and outside of these regimes the probability of an spurious detection is larger. While this should not be technically possible, it is due because these two probabilities come from two different finite sets of realisations, and thus they have an intrinsic variance. If anything, it can be interpreted as the effective probability of a non-spurious detection being null.

Furthermore, the total probability of any detection (regardless of the kind of 2PCF and the binning) has also been computed, considering that any realisation where one or more bins in any 2PCF had SNR>3 was an effective detection. These results are summarised in the two upper rows of Table \ref{tab:probability_uncorrelated}; the probability of a real detection is simply the difference between the probability of any detection (percentage of the 1000 realisations that yielded a detection) an the probability of a spurious detection (percentage of the 1000 realisations with inverted Ly$\alpha$ emission where a detection happened). With this approach, we assume that the probability of a real detection of Ly$\alpha$ cross-correlation and a spurious one are independent processes.
 
 \begin{table}
\centering
	\caption{Probability of any detection for simulated cross-correlations PAUS-eBOSS and PAUS-DESI, using the uncorrelated PAUS noise extrapolation.}
 	\label{tab:probability_uncorrelated}
 	\begin{tabular}{rccc}
 		\hline
 		Surveys & Any detection & Spurious & Real\\
 		\hline
 		PAUS-eBOSS  & 18.7\% & 16.9\% & 1.8\%\\
 		\hline
        PAUS-DESI & 24.7\% & 20.2\% & 4.5\%\\
        \hline 		
        PAUS deep-DESI  & 39.5\% & 24.2\% & 15.3\%\\
 		\hline
        PAUS extended-DESI & 36.8\% & 27.8\% & 9.0\%\\
        \hline
 	\end{tabular}
 \end{table}

When considering these results, it is important to take into account that in this preliminary study no weightings to improve SNR of the estimator in Eq. \ref{eq:corr_estimator} have been considered, and only a uniform binning have been applied for the 2PCFs computation. Nevertheless, the detection probabilities are still very small, with the probability of an spurious detection being far higher than an actual one in both cases; it is safe to assume that any statistical approach to increase SNR is unlikely to yield significantly better results.

\subsubsection{Hypothetical cases: PAUS deep, PAUS extended}\label{sec:hypothetical_detection_probability_uncorrelated}

In addition to the PAUS-eBOSS and PAUS-DESI simulations, two hypothetical cases have also been considered: PAUS deep, a survey with the same field coverage, but complete up to a magnitude deeper ($i_{\rm AB}<24$), and PAUS extended, with the same exposure time as current PAUS, but a larger angular area of 225 deg$^2$. These hypothetical PAUS cases have only been cross-correlated with the DESI simulation, since eBOSS would be rendered obsolete by DESI before such hypothetical surveys could be finished.

PAUS deep has been simulated analogously to PAUS, with the sole difference being the instrumental noise, now reduced by a factor of $\sqrt{6}$, as displayed in Table \ref{tab:sigma_noise_abs} ($\sigma_{\rm 18exp\, abs}$). Regarding PAUS extended, the Ly$\alpha$ emission array has been repeated in a mosaic of 3x3, instead of 2x2, thus yielding an angular coverage of $\sim$225 deg$^2$. To cover this mosaic of Ly$\alpha$ emission, the foregrounds array has been repeated and rotated for the first 4 iterations; after that, it has been mirrored in RA direction and repeated until the 3x3 mosaic has been filled. This gives 8 possible combinations of Ly$\alpha$ emission-foregrounds: the 4 rotations of the foreground array plus the 4 mirrored rotations, which sets a limit on the maximum area we can simulate in this study. In fact, the 3x3 mosaic already has one redundant combination of Ly$\alpha$ emission+foregrounds (since it is composed of 9 realisations). Simulating even larger areas would result in largely redundant foregrounds, which would provide too optimistic results given that the same combination of Ly$\alpha$ emission+foregrounds would be sampled multiple times.

\begin{figure}
 	\includegraphics[width=\columnwidth]{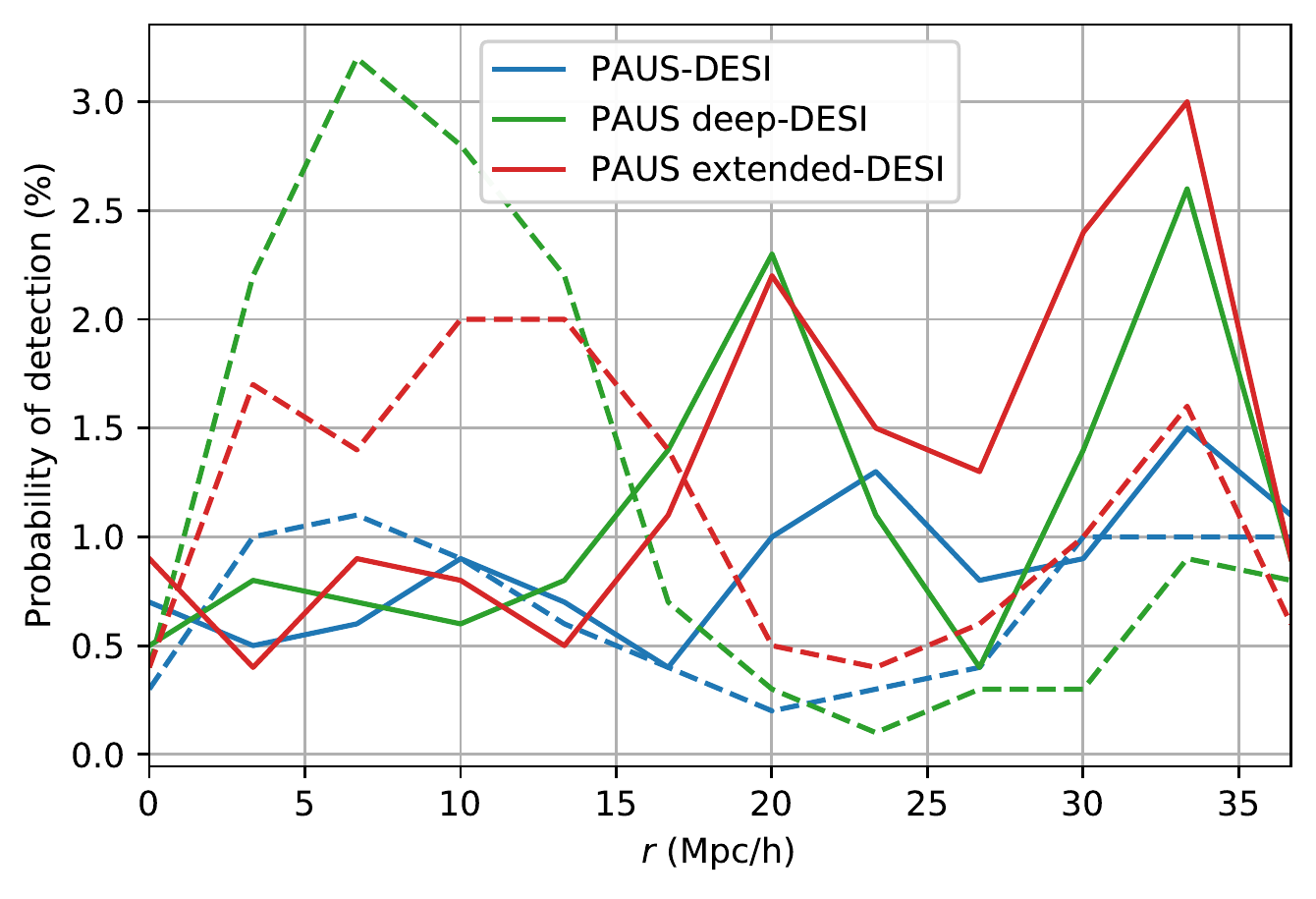}
 	\includegraphics[width=\columnwidth]{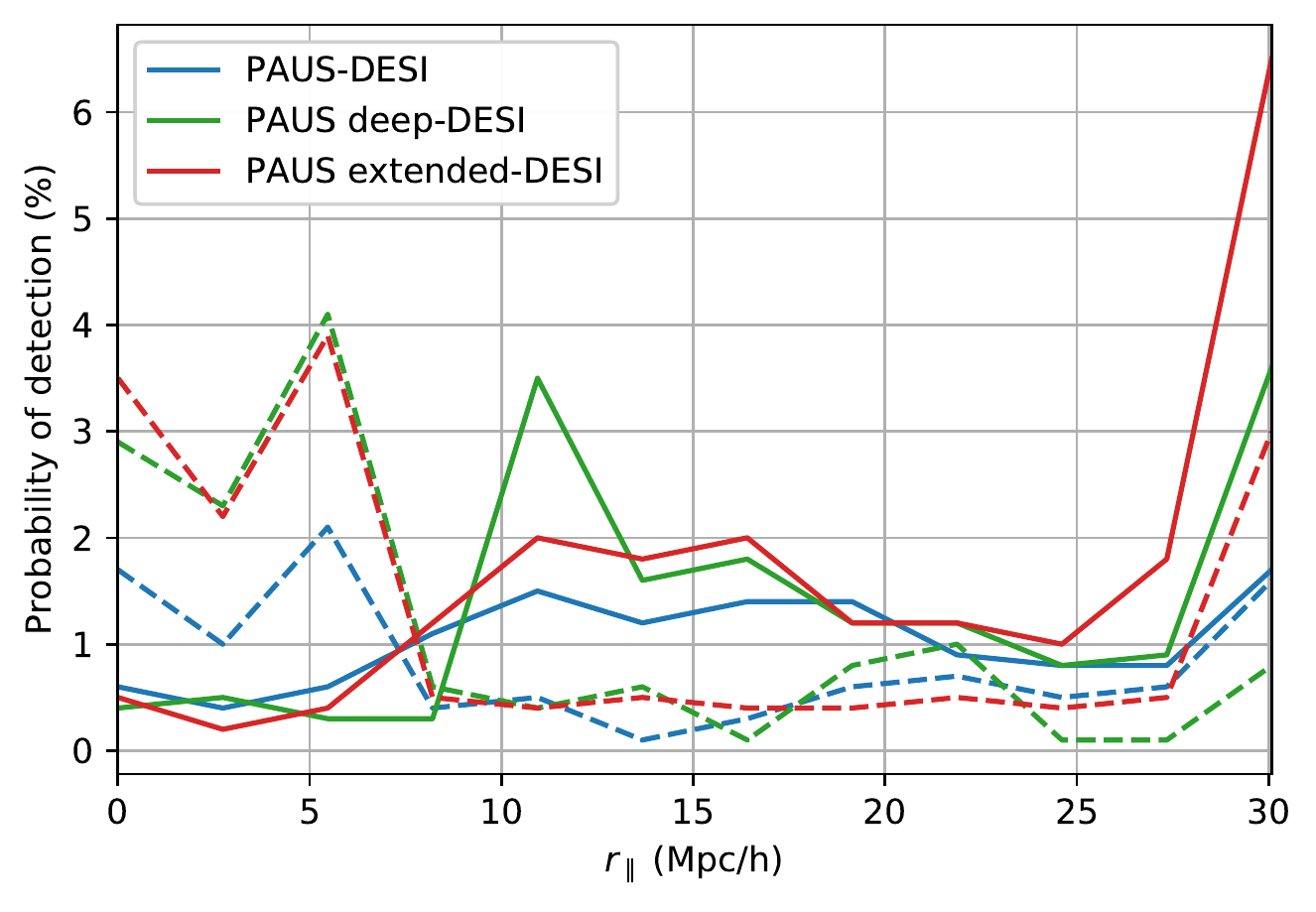}
     	\includegraphics[width=\columnwidth]{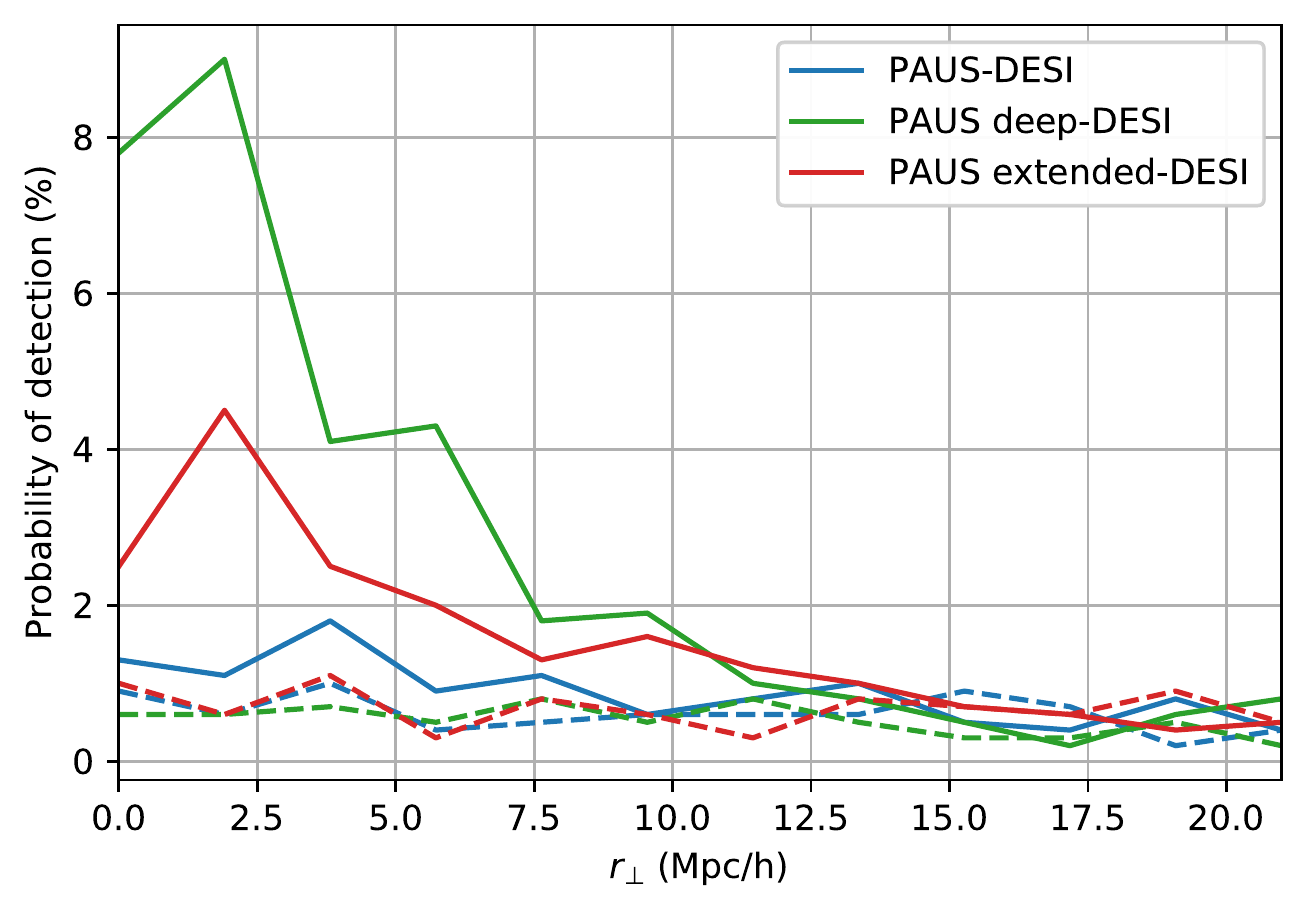}
     \caption{Probability of a detection as a function of distance in the simulated cross-correlation of hypothetical extensions of PAUS and DESI, for 1000 different realisations of instrumental noise+Ly$\alpha$ forest. PAUS deep refers to a survey complete up to $i_{\rm AB}<24$ (exposure time x6), while PAUS extended refers to a total survey area of 225 deg$^2$. Solid line displays the actual probability of any detection, dashed line shows the probability of a spurious detection. \textit{Top panel:} Monopole 2PCF. \textit{Middle panel:} Parallel 2PCF. \textit{Bottom panel:} Perpendicular 2PCF.}
     \label{fig:hypothetical_probability}
\end{figure}

The probability of detection for 1000 realisations of these simulations is shown for the monopole, parallel and perpendicular 2PCFs in Fig. \ref{fig:hypothetical_probability}, together with original PAUS-DESI simulation, while Table \ref{tab:probability_uncorrelated} displays the probability of any detection (two lower columns). The probability of spurious detection has also been computed following the same methodology.

Both cases show a noticeable increase in the probability of any detection, which is now close to 40\%. However, the probability of spurious detections also increases, which diminishes the net gain in the probability of any detection. Overall, spurious detection are more likely, but the probability of a real detection is only smaller by a factor of few (less than 2 for PAUS deep-DESI, and approximately 3 for PAUS extended-DESI).

The same complementary trend is observed in Fig. \ref{fig:hypothetical_probability}, with the perpendicular 2PCF sampling better at scales below 10 Mpc/h, while the monopole and parallel 2PCF have a much higher chance of detection at larger scales. For these last two 2PCFs, PAUS extended seems to provide a much higher increase of probability of detection (an increase by a factor of 2-3) at distances larger than 10 Mpc/h, while the improvement of PAUS deep compared to original PAUS is much smaller. The perpendicular 2PCF at small scales, however, shows similar improvement with either PAUS deep or PAUS extended. PAUS deep, however, seems to perform much better at small scales with the perpendicular 2PCF.

PAUS deep would need 6 times the observation time from current PAUS to observe the same area (going from 3 exposures for each pointing to 18), while PAUS extended only would need 2.25 times the observation time to be carried out (since 225 deg$^2$ are being observed instead of 100 deg$^2$, with the same exposure time per pointing). Since PAUS deep seems to provide better detection probabilities (by a factor of $\sim$1.5), but also requires almost twice the observational time, it is difficult to assess which strategy is more time-efficient for a Ly$\alpha$ IM detection. Nevertheless, it seems clear that increasing exposure time yields better results on small scales, and observing a larger field increases the detection probability at larger scales.

\subsection{Cross-correlation with correlated noise: Probability of detection}\label{sec:detection_probability}

So far in this section, we have only discussed the results for the optimistic case where it is assumed that methods are developed to remove the noise correlation the photometric imaging of PAUS. However, this is not the current case, and while a noticeable reduction in the noise correlation might be achieved (since this is an active area of research in other IM applications, such as 21 cm IM, e.g. \cite{Liu2009}), using the correlated noise is the most realistic approach for now.

Table \ref{tab:probability_correlated} shows the probability detection results, following the same methodology as in \S\ref{sec:detection_probability_uncorrelated}, but applying the correlated noise to the simulation of PAUS images. The probability of a detection greatly decreases in both cases, to the point of being negligible in all cases but PAUS deep-DESI, where it is close to 2\%. In fact, for the PAUS-eBOSS case we actually obtain more spurious detections with the inverted Ly$\alpha$ signal that total detections with the proper cross-correlations; a clear sign of all of them being spurious.

\begin{table}
\centering
	\caption{Probability of any detection for all the considered cases, using the actual correlated PAUS noise.}
 	\label{tab:probability_correlated}
 	\begin{tabular}{rccc}
 		\hline
 		Surveys & Any detection & Spurious & Real\\
        \hline
 		PAUS-eBOSS  & 15.2\% & 17.5\% & 0.0\%\\
 		\hline
        PAUS-DESI & 18.0\% & 17.9\% & 0.1\%\\
 		\hline
 		PAUS deep-DESI  & 18.5\% & 16.8\% & 1.7\%\\
 		\hline
        PAUS extended-DESI & 18.6\% & 18.4\% & 0.2\%\\
        \hline
 	\end{tabular}
 \end{table}

\section{Conclusions}\label{sec:Conclusions}

In this work the possibility of performing Ly$\alpha$ IM by cross-correlation of spectroscopic Ly$\alpha$ forest data with the background of narrow-band images from PAUS has been simulated and evaluated. Ly$\alpha$ forest emission and absorption has been simulated from a hydrodynamic simulation of size 400 Mpc/h designed for the study of the IGM \citep{Cisewski2014, Ozbek2016, Croft2018}. The foregrounds in PAUS images have been simulated from a lightcone mock catalogue made from the Millennium Simulation with the WMAP7 cosmology \citep{Guo2013}, and using the \textsc{galform} (\citealt{Gonzalez-perez2014a}) semi-analytical model. SED templates \citep{Blanton2006} have been fitted using non-negative least squares to the broad-band data of this mock catalogue in order to achieve PAUS spectral resolution for these foregrounds. 

Instrumental noise has been considered for both the simulated PAUS images (measured directly from reduced and stacked PAUS science images) and the Ly$\alpha$ forest spectroscopic data (extracted from \cite{Chabanier2019}). Two different cases for the PAUS instrumental noise have been simulated: one with an optimistic uncorrelated noise extrapolation (assuming that noise correlation is mitigated in future work), and another with the current noise levels directly measured, which show a clear correlation at the pixel size of our simulation.

Furthermore, the theoretical 2PCFs (monopole, parallel and perpendicular correlations) have been computed with the derivation shown in \citet{Gazta2009} from the matter power spectrum, obtained using \textsc{camb} \citep{Lewis2000}. The smoothing of these theoretical 2PCFs due to the large redshift bins for Ly$\alpha$ emission in PAUS narrow-band images has been simulated, and the biases and RSDs of both Ly$\alpha$ emission and absorption have been measured by comparing the theoretical monopole 2PCF to the correlation of the Ly$\alpha$ absorption and emission arrays, using the same spatial binning as the PAUS-DESI cross-correlation.

The simulated cross-correlations without foregrounds or instrumental noise show that, despite the redshift smoothing of Ly$\alpha$ emission in PAUS images, and the limited fraction of space sampled by Ly$\alpha$ forest data, the theoretical monopole 2PCF can be recovered, and the bias of both Ly$\alpha$ emission and absorption can be measured. This shows the validity of this technique in an ideal case to both place constraints on the 2PCF and the bias of the extended Ly$\alpha$ emission or the Ly$\alpha$ forest.

Nevertheless, a bias has been identified in the cross-correlation estimator when cross-correlating fields with noise with mean larger than zero (such as the foregrounds and the instrumental noise for this case). This noise bias, while not affecting the SNR, should be taken into account if constraints such as the Ly$\alpha$ emission bias or the Ly$\alpha$ mean luminosity are to be derived from cross-correlation. A constrained model of the foregrounds and other noise sources average values would be needed; conversely, assuming a known bias and expected Ly$\alpha$ luminosity this same cross-correlation could be used to place constraints on foregrounds emission.

When the cross-correlation is run with the instrumental noise and foregrounds in PAUS images, SNR greatly decreases, up to the point where not all realisations yield a detection. A realisation of this cross-correlation contains three stochastic elements: the instrumental noise of PAUS and the Ly$\alpha$ forest, derived from a random Gaussian distribution (although negligible in the second case), and the positions of the quasars, drawn from the quasar redshift distribution of eBOSS/DESI. Fixing one of these stochastic elements does not provide consistent SNR either, so the probability of a detection (i.e., the cross-correlation reaching a certain SNR threshold) has been evaluated using a purely frequentist approach.

In order to evaluate the probability of a detection, 1000 realisations of the simulated cross-correlations have been carried out with different realisations of both instrumental noise and quasar positions, and for each one the monopole, parallel and perpendicular 2PCFs have been computed for 12 uniform distance bins. Moreover, another 1000 realisations have been computed with mirrored Ly$\alpha$ emission in PAUS images, to determine the probability of spurious detections.

Considering a detection threshold of SNR>3, and under the uncorrelated PAUS noise assumption, Ly$\alpha$ emission has been detected in only 1.8\% of PAUS-eBOSS simulations and 4.5\% of PAUS-DESI simulations. These percentages increase to 15.3\% and 9.0\% with two hypothetical PAUS extensions: PAUS deep (going up to $i_{\rm AB}<24$ instead of $i_{\rm AB}<23$), and PAUS extended (observing 225 deg$^2$ instead of 100 deg$^2$). Nevertheless, in all cases the probability of a spurious detection is higher, and when including the correlated PAUS noise instead, the higher probability of a real detection (PAUS deep-DESI), is just 1.7\%. These results clearly show that, even if noise correlation was to be mitigated and PAUS observation time extended, the cross-correlation of the images background with Ly$\alpha$ forest data is unlikely to yield a detection, and if such a detection happens the most likely scenario is that it is spurious.

Despite these negative results, some valuable conclusions can still be extracted. First, the perpendicular and parallel 2PCF show complementary behaviours: the former has relatively high detection probabilities at scales up to 10 Mpc/h, while the latter displays a non-negligible probability of detection at scales larger than 10 Mpc/h. These different trends are due to the smoothing of the 2PCF in redshift direction, which affects far more the parallel 2PCF than its perpendicular counterpart. Second, this smoothing effect has been properly modelled and recovered when comparing the noiseless correlation to theory, so larger smoothing lengths can be accounted for, and the scales where they will maximize the detection probability, and thus the SNR, can also be predicted.

These two results point out to the fact that broad-band photometric surveys, with angular coverages one or even two orders of magnitude larger than PAUS, such as DES \cite{Abbott2018} or SDSS \citep{Ahumada2020} may be more promising for Ly$\alpha$ IM. This is because their main drawback compared to narrow-band surveys (redshift smoothing) has been properly modelled and reproduced, and increasing survey area has been shown to be an effective strategy to increment the detection probability.

\section*{Acknowledgements}

The PAU Survey is partially supported by MINECO under grants CSD2007-00060, AYA2015-71825,  ESP2017-89838, PGC2018-094773, PGC2018-102021, SEV-2016-0588, SEV-2016-0597, MDM-2015-0509 and Juan de la Cierva fellowship and LACEGAL and EWC Marie Sklodowska-Curie grant No 734374 and no.776247 with ERDF funds from the EU  Horizon 2020 Programme, some of which include ERDF funds from the European Union. IEEC and IFAE are partially funded by the CERCA and Beatriu de Pinos program of the Generalitat de Catalunya. Funding for PAUS has also been provided by Durham University (via the ERC StG DEGAS-259586), ETH Zurich, Leiden University (via ERC StG ADULT-279396 and Netherlands Organisation for Scientific Research (NWO) Vici grant 639.043.512), University College London and from the European Union's Horizon 2020 research and innovation programme under the grant agreement No 776247 EWC. The PAU data center is hosted by the Port d'Informaci\'o Cient\'ifica (PIC), maintained through a collaboration of CIEMAT and IFAE, with additional support from Universitat Aut\`onoma de Barcelona and ERDF. We acknowledge the PIC services department team for their support and fruitful discussions. RACC was supported by  NASA NNX17AK56G, NASA ATP 80NSSC18K101, NSF AST-1614853. and NSF AST-1615940.

\section*{Data Availability Statement}

The data underlying this article will be shared on reasonable request to the corresponding author.




\bibliographystyle{mnras}
\bibliography{bibliography} 







\bsp	
\label{lastpage}
\end{document}